\documentclass[aip,jmp,amsmath,amssymb,
reprint,%
]{revtex4-1}

\usepackage{graphicx,amsmath,amssymb}% Include figure files
\usepackage{dcolumn}% Align table columns on decimal point
\usepackage{bm}% bold math

\usepackage[geometry]{ifsym}
\usepackage[usenames]{color}

\newcommand{\bse}{\begin{subequations}}
\newcommand{\ese}{\end{subequations}}
\newcommand{\be}{\begin{equation}}
\newcommand{\ee}{\end{equation}}
\newcommand{\bfig}{\begin{figure}}
\newcommand{\efig}{\end{figure}}
\newcommand{\bc}{\begin{center}}
\newcommand{\ec}{\end{center}}
\newcommand{\btab}{\begin{tabular}}
\newcommand{\etab}{\end{tabular}}
\newcommand{\dr}{\partial}

\let\oldepsilon\epsilon
\let\epsilon\varepsilon
\let\varepsilon\oldepsilon

\newcommand{\dd}{{\rm d}}
\newcommand{\di}{\partial_i}
\newcommand{\djj}{\partial_j}

\newcommand{\dt}{\partial_t}

\begin{document}

\title{Numerical simulation of turbulent sediment transport, from bed load to saltation.}

\author{Orencio Dur\'an}
	\altaffiliation[present address: ]{Univ N Carolina, Dept Geol Sci, 104 South Rd, Mitchell Hall, Campus Box 3315, Chapel Hill, NC 27515 USA.}
\author{Bruno Andreotti}
\author{Philippe Claudin}
\affiliation{Laboratoire de Physique et M\'ecanique des Milieux H\'et\'erog\`enes,\\
PMMH UMR 7636 ESPCI -- CNRS -- Univ. Paris-Diderot -- Univ. P.M. Curie,\\
10 rue Vauquelin, 75005 Paris, France}
 
%\pacs{45.70.Mg,47.55.Kf,92.40.Gc,47.27.-i}

\date{\today} 

\begin{abstract}
Sediment transport is studied as a function of the grain to fluid density ratio using two phase numerical simulations based on a discrete element method (DEM) for particles coupled to a continuum Reynolds averaged description of hydrodynamics. At a density ratio close to unity (typically under water), vertical velocities are so small that sediment transport occurs in a thin layer at the surface of the static bed, and is called bed load. Steady, or `saturated' transport is reached when the fluid borne shear stress at the interface between the mobile grains and the static grains is reduced to its threshold value. The number of grains transported per unit surface is therefore limited by the flux of horizontal momentum towards the surface. However, the fluid velocity in the transport layer remains almost undisturbed so that the mean grain velocity scales with the shear velocity $u_*$. At large density ratio (typically in air), the vertical velocities are large enough to make the transport layer wide and dilute. Sediment transport is then called saltation. In this case, particles are able to eject others when they collide with the granular bed, a process called splash. The number of grains transported per unit surface is selected by the balance between erosion and deposition and saturation is reached when one grain is statistically replaced by exactly one grain after a collision, which has the consequence that the mean grain velocity remains independent of $u_*$. The influence of the density ratio is systematically studied to reveal the transition between these two transport regimes. Based on the mechanisms identified in the steady case, we discuss the transient of saturation of sediment transport and in particular the saturation time and length. Finally, we investigate the exchange of particles between the mobile and static phases and we determine the exchange time of particles.
\end{abstract}

\maketitle

%\tableofcontents

%%_________________________________________________________
\section{Introduction}
After the pioneering works of Richardson \cite{R26}, Rouse \cite{R36} and Vanoni \cite{V46}, transport and dispersion of impurities suspended in turbulent flows, such as sand grains, dust, bubbles or droplets, have received a renewed interest in the last decade, both from the fundamental point of view \cite{CST98,BBCLMT07,TB09} and  for its applications to planetology \cite{PL01}, cloud physics \cite{KS01} and geomorphology. In the later case, sediment may be entrained, transported and deposited by water flow or by wind. Then, gravity cannot be neglected as transport usually takes place in a turbulent boundary layer bounded by an erodible granular bed. Moreover, transported particles are not passively advected by the flow: they induce a negative feedback, which eventually limits the erosion of the granular bed, leading to a steady state in which erosion and deposition balance each other.

In such a homogeneous and steady situation, the fluid flow can be characterized by an unique quantity: the shear velocity $u_*$, defined by the wall shear stress. The flux  of sediments transported by the flow, called the saturated flux and noted $q_{\rm sat}$, is an increasing function of $u_*$ whatever the nature of the fluid. For aeolian transport, there has been great effort to obtain the relation $q_{\rm sat}(u_*)$ experimentally \cite{CM39,B41,Z53,W64, ST74,N78,JW79,W79,WRS82,GBW96,IR99} using both wind tunnels and atmospheric flows in the field, numerically \cite{AH86,AH88,AH91,W90,KR09} and theoretically \cite{B41,K51,O64,K76,LL78,UH87,S91,SKH01}. Similarly, several expressions for subaqueous bed-load have been proposed, \cite{MM48,E50,B56,Y63,R98,CL05,WP06,LMC10}. Most models are based on the same dynamical mechanisms and differ only by the approximation used to compute the particle trajectories \cite{B73,vR84,EF76,WS89,SK92}. For this reason, experiments have been performed to determine the saltating motion of individual particles under water \cite{LvB76,AF77,NGA94,LH94,CLDZ07} and in air \cite{MWR92,NHB93,FS97,ZWHZLD01}.

Despite this wide literature, some fundamental aspects of sediment transport are still only partly understood. For instance, the dynamical mechanisms limiting sediment transport, in particular the role of the bed disorder \cite{C06} and turbulent fluctuations \cite{C98,B04,VBSA04,BS05,MASS06,B08,LBDVC09}, remain matter of discussion. Also, derivations of transport laws have a strong empirical or semi-empirical basis, thus lacking more physics-related inputs. Here we investigate the properties of sediment transport using a novel numerical description of particle-laden flows, using two-phase numerical simulations based on a discrete element method (DEM) for particles coupled to a continuum Reynolds averaged description of hydrodynamics.
In particular, we examine the transition from bed-load to saltation by studying the influence of the grain to fluid density ratio. A similar approach has recently been used to study the onset of aeolian saltation \cite{CPH11}.

In section II, we introduce the equations of motion for the grains as well as the equations of hydrodynamics, emphasizing the coupling between the two. Then, in section III, we detail the characteristics of saturated transport in the two limiting cases: bed load (water) and saltation (air). In section IV, we propose an interpretation of the simulations based on simple transport models. We then use these transport descriptions to derive and discuss out-of-equilibrium transport and in particular the saturation length and time (section V). We contrast this time with the `exchange time', which characterizes the diffusion of particles through the static/mobile interface. Finally, conclusions are outlined in the last section.

%%_________________________________________________________
\section{Transport model}

\subsection{Key ideas}
We wish to model the transport of non-cohesive grains by a flow, under gravity. Although a continuum `two phase' (grains and fluid) modeling \cite{OAG09a} is very appealing, it is problematic by several aspects. (i) It postulates that particles constitute an Eulerian phase, which means that the particles crossing an arbitrary control volume have almost the same velocity. In a homogeneous steady flow, an Eulerian approach immediately predicts that particles are transported along the direction parallel to the bed and to the flow -- vertical velocities are ignored. However, at least for saltation, they are essential. (ii) Such a continuum approach ignores the discrete and disordered nature of the granular phase. However, these properties are essential close to the transport threshold, below which no grain can be entrained. Such models incorrectly predict the threshold shear velocity and in particular its strong decrease with the grain Reynolds number. To avoid these issues, we use here a discrete element method for the particles \cite{AH86,AH88,AH91}.

Resolving hydrodynamics around grains is technically feasible only if the size of the domain (the number of grains) and the time over which the simulation is run are very small. The idea introduced here is thus to use a continuum description of hydrodynamics, averaged at a scale larger than the grain size. This means that the feedback of the particles on the flow is treated in the mean field manner.

This method allows one to perform very long numerical simulations (typically $1000\,\sqrt{d/g}$), using a (quasi) 2D large spatial domain (typically $15000$ spherical grains in a $xyz$ box of respective dimensions $1000\,d \times 1\,d \times 1000\,d$), while keeping the complexity of the granular phase. Periodic boundary conditions are used in the $x$ (flow) direction. The domain height is large enough as to prevent any grain for reaching the top border. We will now detail the different ingredients of the model. To avoid the formation of ordered structures in the grain packing, we have used a slightly polydisperse sample (the maximum grain diameter is about $20\%$ higher than the mean). For the sake of simplicity, we only give here the equations for the strictly monodisperse case (grains of diameter $d$). 

%%_________________________________________________________
\subsection{Forces on particles}

%%_________________________________________________________
\subsubsection{Equations of motion}
The grains have a spherical shape and are described by their position vector $\vec r$, velocity $\vec u$ and angular velocity $\vec \omega$. A given grain labelled $p$ inside a fluid obeys the equations of motion,
\begin{eqnarray}
	\label{eq:motion}
m\frac{\dd \vec u^p}{\dd t} &=& m\,\vec g + \sum_{q} \vec f^{p,q} + \vec f^p_{\rm fluid}\nonumber \\
I \frac{\dd \vec\omega^p}{\dd t} &=& \frac{d}{2} \sum_{q} \vec n^{p,q} \times  \vec f^{p,q} 
\end{eqnarray}
where $\vec g$ is the gravity acceleration, $I = m d ^2/10$ is the moment of inertia of a sphere, $\vec f^{p,q}$ is the  contact force with grain $q$, $\vec n^{p,q}$ is the contact direction, and $\vec f^p_{\rm fluid}$ encodes forces of hydrodynamical origin.

%%_________________________________________________________
\subsubsection{Contact forces}
Following a standard approach for the modeling of contact forces in MD codes, see \cite{CS79,L98,L08,DEMGM11} and references therein, we consider the case where grains in contact are subject to (i) normal repulsion, (ii) tangential friction and (iii) energy dissipation. For simplicity, the normal repulsion is given by a spring-like elastic force -- the results are independent of the spring stiffness, provided it is large enough. The tangential friction is modeled by a tangential elastic force proportional to the relative tangential displacement between the grains. The moment of this force can induce particle rotation. Whenever the tangential exceeds a given fraction of the normal force, defined by a microscopic friction coefficient, the contact `slides' (Coulomb friction law). Finally, energy dissipation at the contact is ensured by adding a damping term to the force, proportional to the relative contact velocity. This term accounts for the restitution coefficient $e$, i.e. the ratio between grain velocities after and before a collision.

%%_________________________________________________________
\subsubsection{Hydrodynamic forces}
For simplicity we assume that the net hydrodynamical force ($\vec f^p_{\rm fluid}$) acting on a grain $p$ due to the presence of the fluid is dominated by the drag and Archimedes forces, $\vec f^p_{\rm drag}$ and $\vec f^p_{\rm Arch}$, respectively: 
\begin{equation}
	\label{eq:fluid}
	\vec f^p_{\rm fluid} = \vec f^p_{\rm drag} + \vec f^p_{\rm Arch}
\end{equation}
The lift force, lubrication forces and the corrections to the drag force (Basset, added-mass, Magnus, etc.) are neglected. Notice that an important consequence of neglecting lubrication forces is the constancy of the grain's restitution coefficient, which in general is function of the grain's Stokes number \cite{GLP02}.  

\paragraph{Drag force:}
We hypothesize here that the drag force exerted by a homogeneous fluid on a moving grain only depends on the difference between the grain velocity $\vec u^p(x,z)$ and the fluid velocity $\vec u(z)$ at grain's height $z$. Introducing the particle Reynolds number $R_u$ based on this fluid-particle velocity difference $R_u = |\vec u - \vec u^p| d/\nu$, the drag force can be written under the form
\begin{equation}
	\label{eq:drag}
\vec f^p_{\rm drag} = \frac{\pi}{8} \rho_f d^2 C_d(R_u) |\vec u - \vec u^p| (\vec u - \vec u^p)
\end{equation}
where $C_d(R_u)$ is the drag coefficient and $\rho_f$ is the density of the fluid. We use the following convenient phenomenological approximation\cite{FC04}:
\begin{equation}
C_d(R_u) = \left(\sqrt{C^{\infty}_d} + \sqrt{R_u^c/R_u} \right)^2
\label{CdofR}
\end{equation}
where $C^{\infty}_d \simeq 0.5$, is the drag coefficient of the grain in the turbulent limit ($R_u \rightarrow \infty$), and $R_u^c \simeq 24$ is the transitional particle Reynolds number above which the drag coefficient becomes almost constant. 

\paragraph{Archimedes force:}
This force results from the stress which would have been exerted on the grain, if the grain had been a fluid. Thus,
\begin{equation}
	\vec f^p_{\rm Arch} = \frac{\pi}{6}d^3 {\rm div} \sigma^f
\end{equation}
where $\frac{\pi}{6} d^3$ is the grain volume and $\sigma^f_{ij} = -p^f\delta_{ij} + \tau^f_{ij}$ is the undisturbed fluid stress tensor (written in terms of the pressure $p^f$ and the shear stress tensor $\tau^f_{ij}$). In first approximation, the stress is evaluated at the center of the grain.

In the inner region of the boundary layer, where we assume that most of the transport takes place, the Archimedes force can be well approximated by:
\begin{equation}
	\label{eq:arch_x}
	\vec f^p_{\rm Arch} = \frac{\pi}{6}d^3 \left( \dr_z \tau^f_{xz} \vec e_x - \dr_z p^f \vec e_z \right)
\end{equation}
where $\vec e_x$ and $\vec e_z$ are the horizontal and vertical directions, respectively.

%%_________________________________________________________
\subsection{Hydrodynamics}

In the presence of particles occupying a volume fraction $\phi$, the hydrodynamics is described by the two-phase flow Reynolds averaged Navier-Stokes equations:
\begin{equation}
\rho_f (1 - \phi) {\rm D}_t u_i= -\di p^f+ \rho_f (1-\phi) g_i + \djj \tau^f_{ij} - F_i
\end{equation}
where ${\rm D}_t u_i \equiv \dt u_i + u_j \djj u_i$ denote the fluid inertia. $\tau^f_{ij}$ is the total shear stress tensor resulting both from viscous diffusion of momentum (viscous stress) and transport of momentum by turbulent fluctuations (Reynolds stress). $\vec F$ is the body force exerted by the grains on the fluid. It reflects the velocity fluctuations induced by a moving grain. As we focus in this paper on steady homogeneous sediment transport, we hypothesize that the influence of a given grain remains localized in a thin horizontal region and that the typical horizontal distance over which the flow is disturbed is comparable to the distance between moving grains. $\vec F(z)$ can then be obtained by averaging the hydrodynamical force $\vec f_{{\rm fluid}}^p$ acting on all the grains moving around altitude $z$, in a horizontal layer of area $A$ and thickness $\dd z$:
\begin{equation}
\label{eq:F}
\vec F(z)=\frac{1}{A\dd z} \left < \, \sum_{p\in \{z;z+\dd z\}} \vec f_{{\rm fluid}}^p \right >.
\end{equation}
Here, we take for $A$ the total horizontal extent of the domain (i.e. $1000 d \times 1 d$). Notice that the number of grains given by the condition $p\in \{z;z+\dd z\}$ is also a differential quantity. The symbols $\langle . \rangle$ denote ensemble averaging. In order to gain statistics, we make use of the steady character of the studied situation, and also use time averaging. For simplicity, we note $\tau^f=\tau^f_{xz}$ the fluid shear stress, and $u = u_x$ for the fluid horizontal velocity. After substituting the drag and Archimedes forces explicitly, $\vec F(z)$ can be rewritten as,
\begin{equation}
	\label{eq:F_1}
	\vec F(z) = \phi \left < \vec F_{{\rm drag}} \right > + \phi \left( \dr_z \tau^f \vec e_x - \dr_z p^f \vec e_z \right).
\end{equation}
where the grain's volume fraction $\phi$ is defined as
\begin{equation}
	\phi(z) = \frac{1}{A \dd z} \sum_{p\in \{z;z+\dd z\}} \frac{\pi}{6} d^3
\end{equation}
and
\begin{equation}
	\label{eq:ff}
	\left < \vec F_{{\rm drag}} \right > (z) = \left < \, \sum_{p\in \{z;z+\dd z\}} \vec f_{{\rm drag}}^p \right > / \sum_{p\in \{z;z+\dd z\}} \frac{\pi}{6} d^3
\end{equation}
is the average drag force acting on grains at height $z$ per unit grain's volume.

In the inner region of the turbulent boundary layer, both the fluid inertia and the horizontal stress gradients can be neglected, and the vertical component of the Reynolds equation becomes 
\bse
\begin{eqnarray}
		 		0 		& = & - \dr_z p^f - \rho_f (1-\phi) g - F_z \\
\mbox{or} \qquad 	\dr_z p^f	& = & - \rho_f g - \frac{\phi}{1-\phi} \left < F_{{\rm drag}, \, z} \right >
\end{eqnarray}
\ese
Under the assumption of steady and homogeneous sediment transport, the contribution of grains' vertical drag to the momentum balance is negligible and the vertical balance reduces to the hydrostatic pressure: 
\begin{equation}
	\dr_z p^f = - \rho_f g
\end{equation}
and thus the Archimedes force (\ref{eq:arch_x}) simplifies to
\begin{equation}
	\label{eq:arch}
	\vec f_{\rm Arch} = \frac{\pi}{6}d^3 \left( \dr_z \tau^f \vec e_x - \rho_f\vec g \right)
\end{equation}
which reduces to the buoyancy force for a static fluid ($\dr_z \tau^f = 0$).

Furthermore, after neglecting inertia and horizontal stress gradients, the horizontal component of the Reynolds equation becomes
\bse
\begin{eqnarray}
	\label{eq:dtauf}
	\dr_z \tau^f & = & F_x
\end{eqnarray}
\ese
which, with the use of the $x$-component of Eq.~\ref{eq:F_1}, can be rewritten as
\begin{equation}
	\label{eq:dtauf2}
	\dr_z \tau^f =  \frac{\phi}{1-\phi} \left < F_{{\rm drag}, \, x} \right >
\end{equation} 
The prefactor $(1-\phi)^{-1}$ accounts for the increasing role of the Archimedes force at the bed, where the volume fraction reaches its maximum value $\phi_b$. Notice however that this increase is balanced by the decrease of the average drag force as the fluid velocity approach zero inside the bed.

For simplicity, in what follows we retain the grain's horizontal feedback term $F_x$, including Archimedes' force, and integrate (\ref{eq:dtauf}) as,
\begin{equation}
	\label{eq:taufp}
	\tau^f(z) = \rho_f u_*^2- \tau^p(z)
\end{equation}
where we have introduced the shear velocity $u_*$, defined by the undisturbed (grain free) wall shear stress, and the grain borne shear stress $\tau^p$ defined by
\bse
\begin{eqnarray}
\tau^p(z) & \equiv & \int_z^{\infty} F_x(z') \dd z' \\
& = &  \int_z^{\infty} \frac{\phi}{1-\phi} \left < F_{{\rm drag}, \, x} \right > \dd z'
\label{eq:taup}
\end{eqnarray}
\ese
where we have replaced $\partial_z \tau^f$ by (\ref{eq:dtauf2}). In the integration bounds, $\infty$ means that the integral includes all moving grains. Here we are assuming that grain's motion takes place well inside the boundary layer, such that at the upper limit of the transport layer, where $F_x = 0$, the horizontal shear stress is constant and equals $\rho_f u_*^2$.

In order to relate the fluid borne shear stress to the average fluid velocity field, we adopt a Prandtl-like turbulent closure \cite{P25}. Introducing the turbulent mixing length $\ell$, we write
\begin{equation}
	\label{eq:tauf_u}
\tau^f = \rho_f (\nu + \ell^2 |\dr_z u|)\dr_z u.
\end{equation}
$\nu$ is the viscosity (a constant independent of the volume fraction). As for the mixing length $\ell$, we know it should vanish below some critical Reynolds number $R_c$ and should be equal to the distance to the surface $z$, far above the transport layer. A common phenomenological approach is to express the turbulent mixing length as a function of the Reynolds number and $z$ (see below Eq.~\ref{eq:l2}). However, this involves the definition of an interphase between the static and mobile zones, below which $\ell$ must vanish. To avoid the need of such a somewhat arbitrary definition, we propose instead a differential equation
\begin{equation}
	\label{eq:l}
\partial_z \ell = \kappa \left[1 - \exp{\left(-\sqrt{\frac{1}{R_c}\left(\frac{u \ell}{\nu}\right)}\right)}\right]
\end{equation}
where $\kappa\simeq 0.4$ is von Karman's constant and the dimensionless parameter $R_c = 7$ is determined from an indirect comparison to measurements (see below). The ratio $u \ell/\nu$ is the local Reynolds number based on the mixing length. Note that a function other than the exponential can in principle be used, provided it has the same behavior in $0$ and $-\infty$, although the present choice provides a quantitative agreement with standard data. As discussed below, this formulation allows us to define $\ell$ both inside and above the static granular bed. Note that, although it appears here as a side technical aspect of the present study on sediment transport, Eq.~\ref{eq:l} is in fact an original piece of work, interesting in itself, and which can be used in other contexts.

%_________________________________________________________
\subsection{Solutions for special cases and validation}

Before presenting the steady state properties of the transport model, let us first provide some technical details of the integration of the model and compare some of the results to what is known for two simple cases: the law of the wall for smooth and rough surfaces. We will then briefly discuss the hydrodynamics for stationary transport. 

\subsubsection{Smooth surface} % (fold)
\label{ssub:smooth_surface}

In the absence of grains, $\tau^f = \rho_f u_*^2$ and the flow velocity $u(z)$ is given by (\ref{eq:tauf_u}), with the boundary condition $u(0) = 0$, and some mixing length phenomenological equation. A popular choice for the flow over smooth surfaces is arguably the one suggested by van Driest \cite{D56,P00}
\begin{equation}
	\label{eq:l2}
	\ell = \kappa z \left[ 1 - \exp{\left( -\frac{1}{R_{\rm vD}} \, \frac{z u_*}{\nu} \right)} \right],
\end{equation}
which reproduces well classical experimental results, in particular the transition from the viscous sub-layer where the velocity profile is linear $u(z)/u_* = z u_*/\nu$, to the turbulent region, where the velocity profile is logarithmic $u(z)/u_* = \kappa^{-1} \log{z/z_0}$. In this smooth case, the hydrodynamic roughness length $z_0$ scales on the viscous length: $z_0 \simeq 0.1 \nu/u_*$. Quantitative agreement with this scaling law is obtained with $R_{\rm vD} \simeq 26$.

\begin{figure}[t!]
\includegraphics{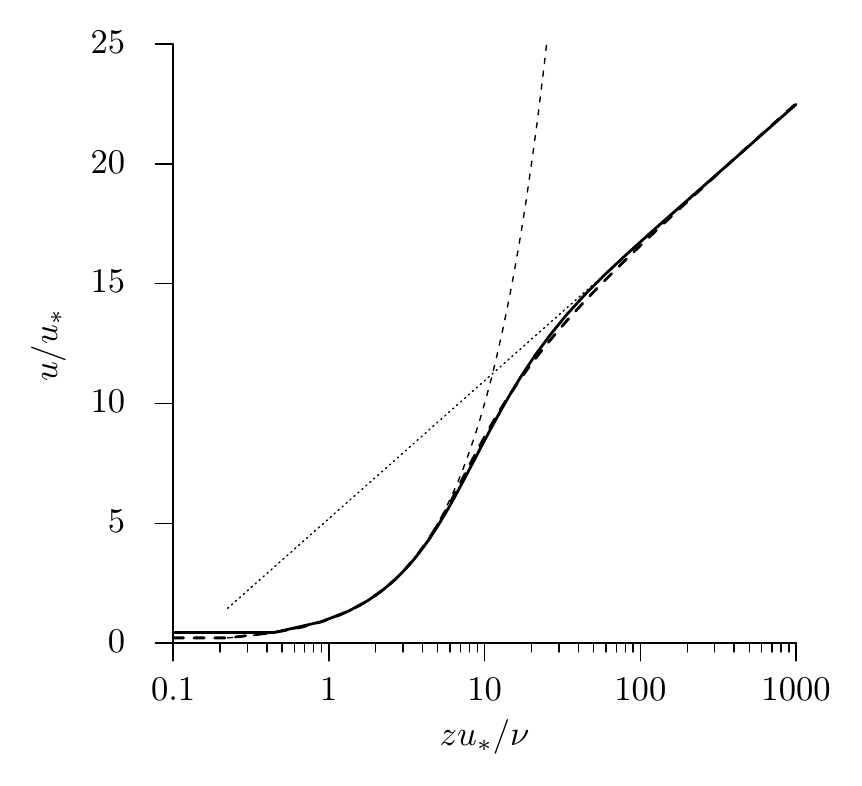}
\caption{Comparison of the normalized vertical velocity profile using the differential mixing length equation proposed here (\ref{eq:l}) with $R_c=7$ (long-dashed line) and the standard van Driest equation (\ref{eq:l2}) with $R_{\rm vD} = 26$ (solid line). The two asymptotic behaviors are also shown: the linear velocity profile at the viscous sub-layer (dashed line) and the logarithmic one far from the surface (dotted line).}
\label{figUwall}
\end{figure}

In contrast to van Driest's mixing length, which can be directly computed for any distance $z$ to the surface, the use of the differential equation (\ref{eq:l}) necessitates to start from a boundary condition at $z=0$. Expanding $\dr_z \ell$ around $u \ell =0$ and substituting the viscous velocity profile $u = z u_*^2/\nu$, we can get the asymptotic behavior of $\ell(z)$ close to the surface and then use it as the boundary condition for $z\rightarrow 0$. The parameter $R_c = 7$ is then fitted to achieve the best comparison of the velocity profile obtained by integration of Eq.~\ref{eq:l} to the one obtained using the van Driest's expression (\ref{eq:l2}). Figure~\ref{figUwall} shows that this comparison is quantitative.

%%%%%%%%%%%%%%%%%%%%%%%%%%%%%%%%%%%
%%%%%%%%%%%%%%%%%%%%%%%%%%%%%%%%%%%

% subsubsection smooth_surface (end)

\subsubsection{Static bed} % (fold)
\label{ssub:static_bed}

The real advantage of Eq.~\ref{eq:l} is evident in the presence of a granular bed. In this case, the interface between the bed and the fluid is not well defined, making the use of Eq.~\ref{eq:l2} difficult. However, as shown below, Eq.~\ref{eq:l} can be integrated from an arbitrary position inside the bed, where the asymptotic solution ($z \rightarrow -\infty$) holds, so that the result of integration is independent of that position.

\paragraph{Asymptotic velocity profile:}
Let us consider a static  ($u^p = 0$) and homogeneous bed with an average volume fraction $\phi_b$. As the flow velocity inside the bed is very small, the average drag force per grain's volume can be approximated by the viscous drag $\left < F_{{\rm drag}, \, x} \right > = 18 \rho_f \nu u/d^2$. Besides, the mixing length is also very small and the fluid borne shear stress is dominated by the viscous diffusion $\tau^f = \rho_f \nu \dr_z u$. From (\ref{eq:dtauf2}) it follows:
\begin{equation}
	\frac{\partial^2 u}{\partial z^2} = \frac{\phi_b}{1-\phi_b} \frac{18}{d^2} u(z)
\end{equation}
The solution of this equation is an exponential decay of the velocity inside the bed with a characteristic `penetration' length 
\begin{equation}
\lambda_b = d \sqrt{(1-\phi_b)/(18\phi_b)} 	
\end{equation}
In our case, we use spherical bed with $\phi_b \simeq 0.6$ and thus $\lambda \simeq 0.2 d$. Integration of the velocity can be started at any arbitrary position $z_s$ well inside the bed (typically 4-5 grains depth is enough). The starting value $u_s$ is found iteratively to ensure that the first integration steps follow the asymptotic behavior
\begin{equation}
	u(z) = u_s \exp{\left( (z - z_s)/\lambda_b \right)}.
\end{equation}

\paragraph{Asymptotic mixing length profile:}
As both the velocity and the mixing length are vanishingly small inside the bed, (\ref{eq:l}) expands as
\begin{equation}
	\dr_z \ell \simeq \kappa\sqrt{u\ell/\nu R_c}	
\end{equation}
which, after substituting the exponential velocity profile, integrates as 
\begin{equation}
	\ell(z) = \frac{\kappa^2\lambda_b^2}{\nu R_c} u(z)
\end{equation}
The proportionality between $\ell$ and $u$ is then used to set the initial value for the mixing length well inside the bed.

%%%%%%%%%%%%%%%%%%%%%%%%%%%%%%%%%%%
\begin{figure}[t!]
\includegraphics{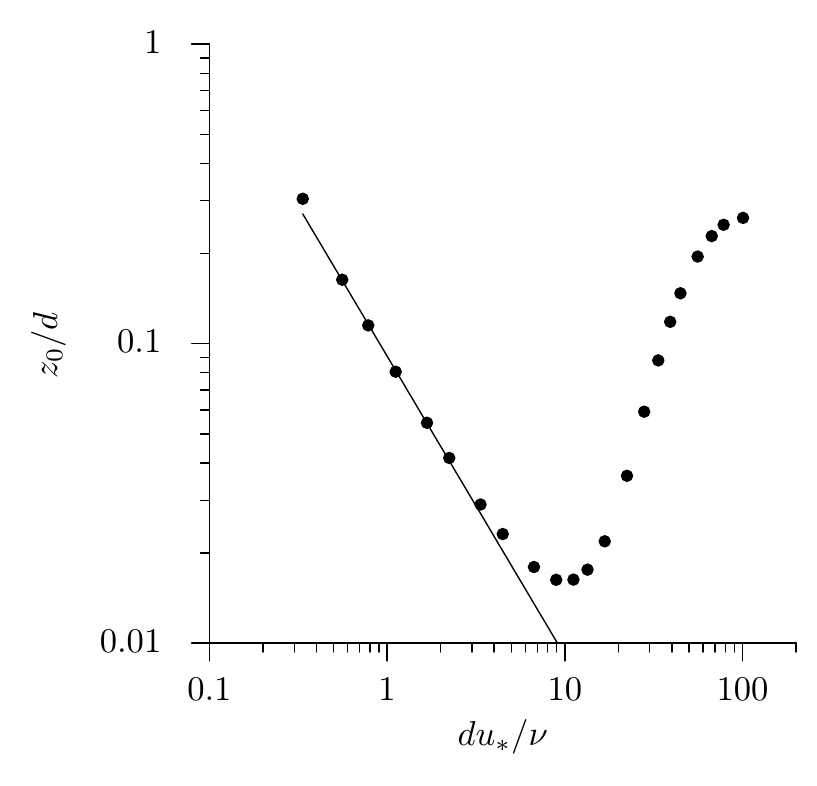}
\caption{Dimensionless roughness length for increasing ratios of grain diameter to viscous length $d u_*/\nu$ (symbols). The prediction for smooth surfaces $z_0 \simeq 0.1 \nu/u_*$ (full line) is also shown for comparison.}
\label{fig_z0}
\end{figure}
%%%%%%%%%%%%%%%%%%%%%%%%%%%%%%%%%%%

\paragraph{Solution for a static bed:}
From velocity profiles, we see two characteristic length scales: the penetration length $\lambda \simeq 0.2 d$ determined by the bed volume fraction and the viscous length $\nu/u_*$ in the viscous sub-layer. For the numerical integration of the velocity we use a fraction of the smallest scale of the two as the mesh size $\dd z$. The mesh is linear inside and just above the bed. Few grains above it, we proceed integration by exponential vertical steps, to adjust better to the logarithmic velocity region. After integration, the flow velocity can be computed at any position by either linear or logarithmically interpolation.

Figure \ref{fig_z0} shows the hydrodynamical roughness length, obtained from the logarithmic velocity profile far from the bed surface, for different ratios of the grain diameter $d$ to the viscous length $\nu/u_*$. There are two distinct regimes: (i) when the viscous length is on the order or higher than the grain diameter, the granular surface behaves as hydrodynamically smooth, with the roughness length proportional to the viscous length $z_0 \propto \nu/u_*$; (ii) when the grain diameter is much larger than the viscous length, there is turbulent dissipation at the surface and the bed becomes hydrodynamically rough with the roughness length seeming to converge to the value around $d/5$. Note that this value of the roughness is larger than what is classically reported from experiments ($z_0 \simeq d/30$ in \cite{B41}; $z_0 \simeq d/24$ in \cite{SG00}; $z_0 \simeq d/10$ in \cite{K74}) we don't expect quantitative agreement working with a two-dimensional model, and model parameters could be more refined for this purpose.

% subsubsection static_bed (end)

\subsubsection{Mobile bed} % (fold)
\label{ssub:mobile_bed}

\begin{figure}[t!]
\includegraphics{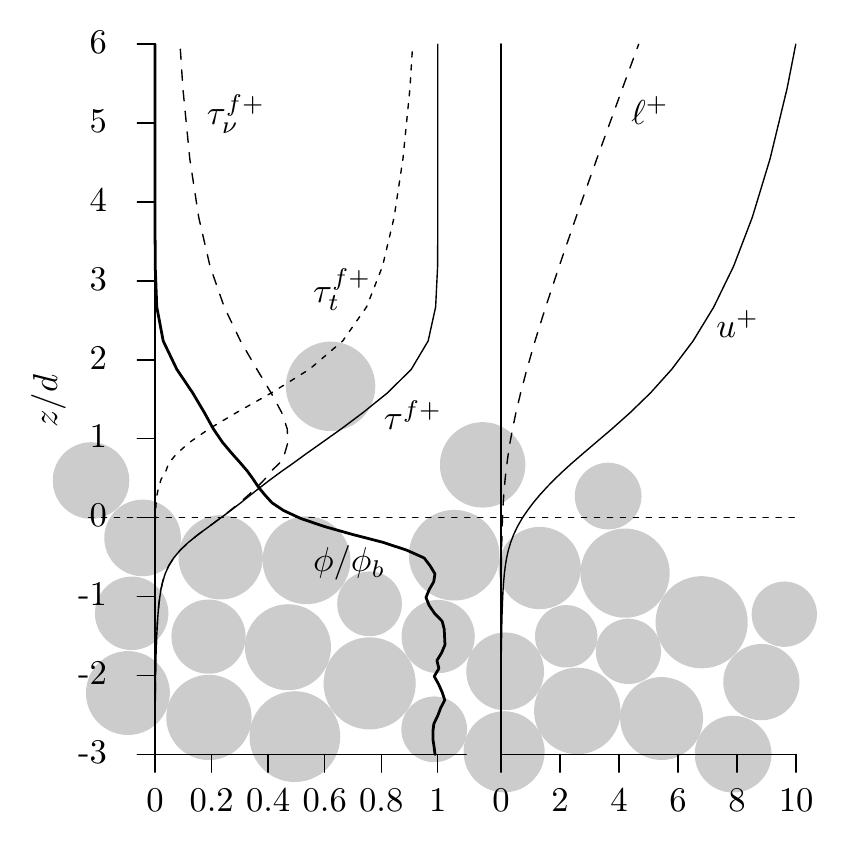}
\caption{Vertical profiles of the rescaled volume fraction $\phi/\phi_b$, flow velocity $u^+ = u/u_*$, mixing length $\ell^+ = \ell/d$, fluid borne shear stress $\tau^{f+} = \tau^f/\rho_f u_*^2$, viscous shear stress $\tau^{f+}_{\nu} = \nu \dr_z u/u_*^2$ and turbulent shear stress $\tau^{f+}_t = (\ell \dr_z u)^2/u_*^2$ (by definition $\tau^{f+} = \tau^{f+}_{\nu} + \tau^{f+}_t$). The reference height $z=0$ is set at the altitude such that $\phi = \phi_b/2$.}
\label{fig_hydro}
\end{figure}

When grains of the bed move, their motion is confined to the surface and the fluid velocity profile is still exponential inside the bed. Therefore, we can use the same procedure to integrate the equations as for a static bed above. As an example, Fig.~\ref{fig_hydro} shows the hydrodynamic profiles for the steady state of a transport simulation with density ratio $\rho_p/\rho_f = 2$. In what follows, as a convention for all vertical profiles, the reference height $z=0$ is by definition the height at which the volume fraction equals half the bed one, $\phi(0) \equiv \phi_b/2$. Notice this definition is not related to grain motion, and finite grains velocity are effectively observed for negative values of $z$. 

Once the grains are able to move, a transport layer quickly develops, accompanied by a strong negative feedback on the flow, and the subsequent reduction of the fluid borne shear stress $\tau^f$ towards zero inside the bed (Fig.~\ref{fig_hydro}). As expected, close to the bed (typically $z < d$) the viscous shear $\tau^f_{\nu} = \rho_f\nu \dr_z u$ is dominant as the mixing length is very small (about $0.01d$). Notice that all profiles are smooth and continuous across the bed surface: (i) in the static bed ($\phi/\phi_b \simeq 1$) there is no turbulence and the balance of viscous drag and viscous momentum dissipation leads to the exponential decay of the flow velocity; (ii) at bed surface ($\phi/\phi_b \lesssim 1$) grains within the transport layer are accelerated until the momentum dissipated due to collisions balance the momentum extracted to the flow; and (iii) at the undisturbed boundary layer ($\phi/\phi_b \rightarrow 0$) the fluid shear stress is constant and the flow momentum is dissipated through viscosity and mostly turbulent mixing (Fig.~\ref{fig_hydro}).

% subsubsection mobile_bed (end)

%%%%%%%%%%%%%%%%%%%%%%%%%%%%%%%%%%%
%%%%%%%%%%%%%%%%%%%%%%%%%%%%%%%%%%%
%
\begin{table}[t!]
\caption{Units used in the model, expressed in terms of the grain density ($\rho_p$), fluid density ($\rho_f$), gravity ($g$) and mean grain diameter ($d$)}
\begin{ruledtabular}
\btab{lr}
General & \\
\hline
length $l$   & $d$ \\
acceleration & $g$\\
time $t$     & $\sqrt{d / g}$ \\
velocity $v$ & $\sqrt{g d}$ \\
\hline\hline
Particles & \\
\hline
angular velocity $\omega$ & $\sqrt{g / d}$ \\
mass $m$ & $\frac{\pi}{6}\rho_p d^3$ \\
moment of inertia $I$ & $m d^2$ \\
force $f$ & $m g$ \\
contact stiffness $k$ & $m g / d$ \\
damping constant $\gamma$ & $m \sqrt{g / d}$ \\
\hline\hline
Fluid & \\
\hline
shear stress $\tau$ & $ (\rho_p-\rho_f) g d $ \\
\etab
\end{ruledtabular} 
\label{tab:units}
\end{table}

%_________________________________________________________
\subsection{Dimensionless numbers for transport}
We now discuss how to make the model equations dimensionless. Gravity gives the relevant scale for forces. More precisely, it only appears in the grain equation of motion under the form of a buoyancy-free gravity $\left(1-\frac{\rho_f}{\rho_p}\right)g$. The choice of the typical length scale is less obvious. On the one hand, the contact forces and the trapping of particles at the surface of the bed do not depend on the fluid properties: the grain diameter $d$ is thus the relevant length scale for the static grains. On the other hand, one can build a drag length from hydrodynamics, which is the length needed to accelerate a grain to the fluid velocity. This inertial length is proportional to $\frac{\rho_p}{\rho_f}\,d$ and is the relevant length scale for the mobile grains. This means that the density ratio $\rho_p/\rho_f$ cannot be eliminated and is a true dimensionless parameter of the problem. We shall see below that this density ratio is the parameter controlling the transition from bed load to saltation. We have chosen $d$ as a reference length scale, and Table~\ref{tab:units} summarizes all the parameters used in our code. The second control parameter is the shear velocity $u_*$ imposed far from the bed, or equivalently the shear stress $\rho_f u_*^2$. Its dimensionless counterpart is the Shields number \cite{S36}, defined by
\begin{equation}
\Theta=\frac{\rho_f u^2_*}{(\rho_p-\rho_f) g d}\;,
\end{equation}
which encodes the strength of the flow. Making the viscosity non-dimensional, we obtain a grain-based Reynolds number
\begin{equation}
R_e = \frac{d}{\nu} \sqrt{\left(\frac{\rho_p}{\rho_f}-1\right) gd}
\end{equation}
Physically, it determines the hydrodynamic regime at the scale of the grain. The figures presented in this paper are obtained using the same particle Reynolds number $R_e = 10$. This value is sufficiently large to ensure that the grain diameter is about one order of magnitude larger than the viscous length. As a consequence, the flow is fully turbulent at about one or two grain layers above the static grains. 

Dynamics at the scale of the contact between grains is controlled by other numbers: the (here constant) restitution coefficient $e = 0.9$, the friction coefficient $\mu = 0.5$ and the contact duration $t_c=\pi / \sqrt{2 k/m - (\gamma/m)^2}$ (in terms of the contact stiffness $k = 5000\,m g/d$ and the damping constant $\gamma = 4.7\,m \sqrt{g/d}$, see \cite{L98}). We have checked that the values given to these parameters do not qualitatively change the results. 

%%_________________________________________________________
\section{Saturated transport}
%%_________________________________________________________
\subsection{Qualitative results}
Transport equations are integrated until a statistically steady homogeneous state is reached. Since we are primarily interested in the transition from bed load to saltation, we have varied the Shields number within the range $\Theta = 0.003$--$0.5$ (a range which contains the threshold $\Theta_d$, see below) and the density ratio within the range $\rho_p/\rho_f = 2$--$2000$.

\begin{figure}[t!]
\includegraphics{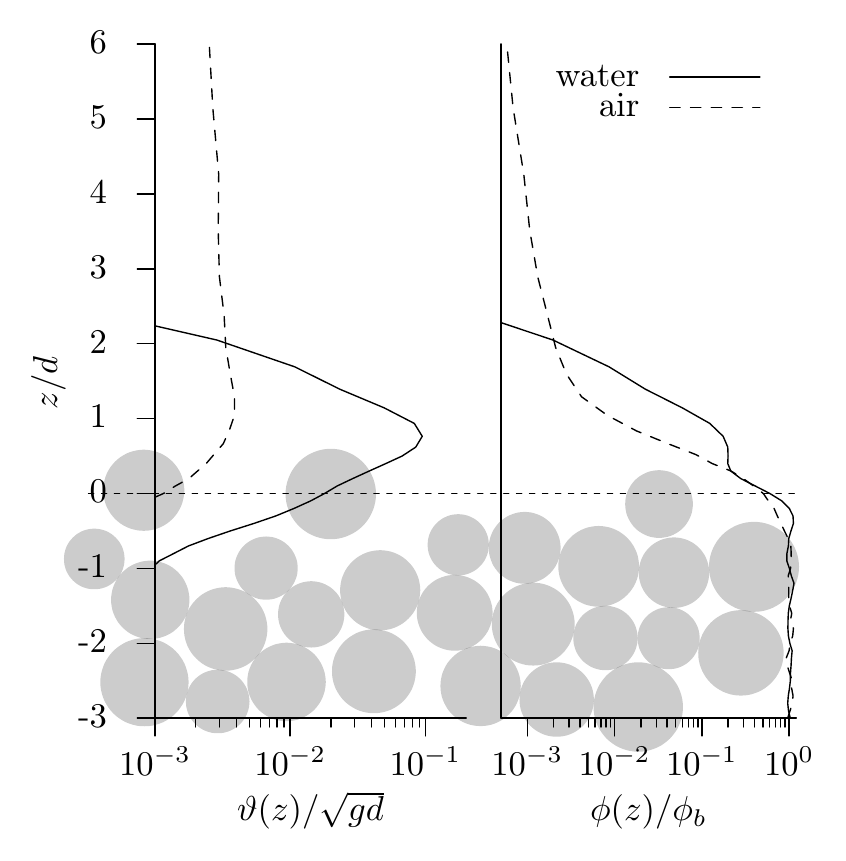}
\caption{Transport profiles: volume flux density $\vartheta(z)$ (left) and volume fraction $\phi(z)$ (right) for water (solid lines) and air (dashed lines).}
\label{fig1}
\end{figure}

Once transport has reached its saturated state, the general picture is as follows: at small density ratios $\rho_p/\rho_f \simeq 2$, which is the typical value under water, the transport is confined at the surface, within a few grain diameters. The dense and thin transport layer is characteristic of the bed load regime. On the contrary, at a large density ratio $\rho_p/\rho_f \simeq 2000$, which is typical of aeolian situation, the transport layer becomes wide and dilute, extending over several tens of grain diameters  (Fig.~\ref{fig1}). This is typical of the saltation regime. Within the very same numerical model, we are thus able to reproduce the basic characteristics of transport at both limits.

%%%%%%%%%%%%%%%%%%%%%%%%%%%%%%%%%%%
%%%%%%%%%%%%%%%%%%%%%%%%%%%%%%%%%%%

%%_________________________________________________________
\subsection{Saturated flux}
\begin{figure}[t!]
\includegraphics{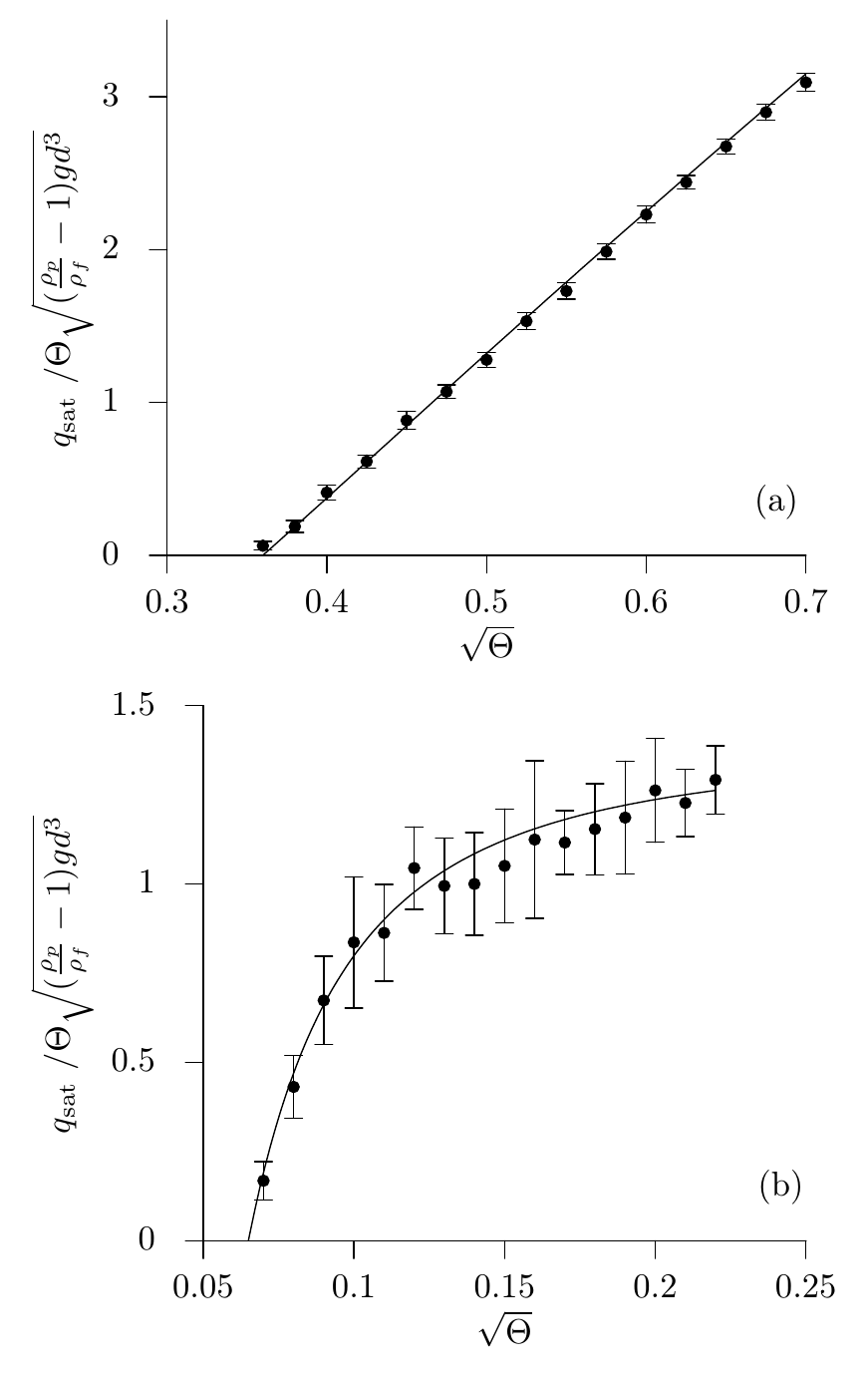}
\caption{Rescaled saturated flux $q_{\rm sat} / \Theta \sqrt{(\rho_p/\rho_f-1) gd^3}$ versus the rescaled shear velocity $\sqrt{\Theta}$ for water (a) and air (b). For air the saturated flux scales asymptotically as $\Theta$ while for water it follows $\Theta^{3/2}$. Full lines are the predictions of the simplified models for bed load (Eq.~\ref{qsatbedload}) and saltation (Eq.~\ref{qsatsaltation}), given in the text.}
\label{figQ}
\end{figure}

Steady and homogeneous sediment transport is basically quantified by the volumetric saturated flux $q_{\rm sat}$, i.e. the volume of the particles (at the bed density) crossing a vertical surface of unit transverse size per unit time. It has the dimension of a squared length per unit time. In the simulations, we compute it as
\begin{equation}
q_{\rm sat} = \frac{1}{A \phi_b} \, \frac{\pi}{6} d^3 \sum_p u^p,
\label{defqsatnum}
\end{equation}
A key issue is the dependence of $q_{\rm sat}$ on the shear velocity or, equivalently, on the Shields number $\Theta$. In order to highlight this dependence, figure~\ref{figQ} shows the saturated flux rescaled by $\Theta$ in both cases (water and air). In agreement with experimental observations~\cite{MM48,E50,B56,Y63,R98,LMC10,RM91,RIR96,IR99,A04,CDOVCJPR09}, we find that $q_{\rm sat}$ scales asymptotically as $\Theta$ (or $u_*^2$) for saltation, while  $q_{\rm sat}$ scales as $\Theta^{3/2}$ (or $u_*^3$) underwater (Fig.~\ref{figQ}). Most models of aeolian transport miss the influence of the negative feedback of transport on the flow. Therefore, they do not give the correct scaling, predicting $q_{\rm sat} \propto u_*^3$. We demonstrate below, in the same numerical model, a fundamental difference between the two transport regimes, which correspond to different underlying dynamical mechanisms.

%%%%%%%%%%%%%%%%%%%%%%%%%%%%%%%%%%%
%%%%%%%%%%%%%%%%%%%%%%%%%%%%%%%%%%%

\begin{figure}[t!]
\includegraphics{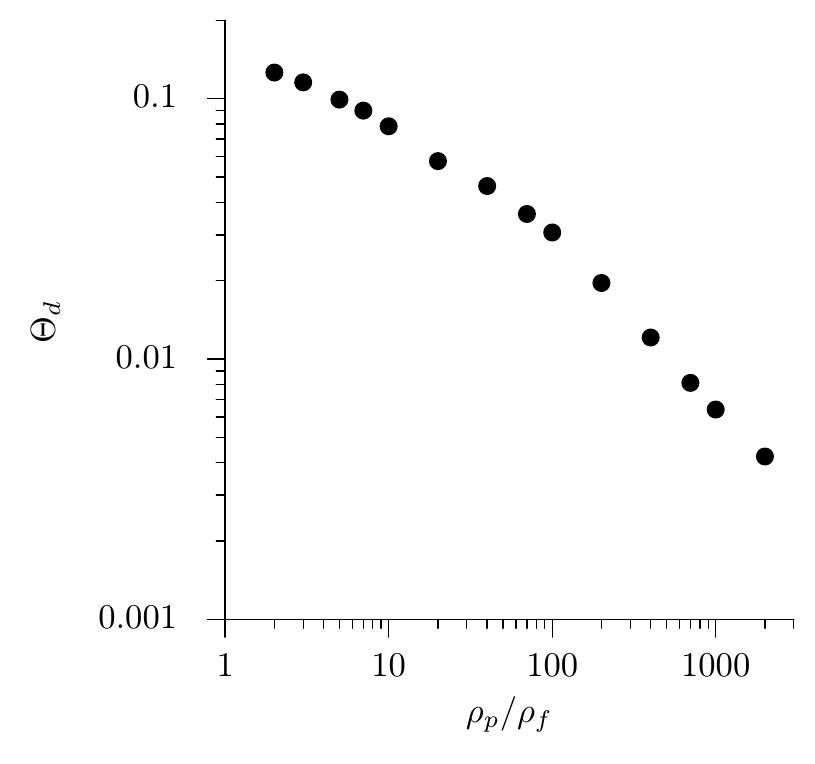}
\caption{Dynamical threshold Shield number $\Theta_d$ as a function of the  density ratio $\rho_p/\rho_f$.}
\label{figUthS}
\end{figure}

Figure~\ref{figQ} reveals the existence of a threshold shear velocity below which the flux vanishes. More precisely, we define the dynamical threshold Shield number $\Theta_d$ from the extrapolation of the saturated flux curve to $0$, which gives in our case $\Theta_d \simeq 0.12$ for water ($\rho_p/\rho_f =2$) and $\Theta_d \simeq 0.004$ for air ($\rho_p/\rho_f =2000$), respectively. These values are consistent with experimental ones within a factor of $2$. Once again, a refined tuning of these values could be achieved by adjusting the model parameter (e.g. $R_c$) and performing 3D simulations. Figure~\ref{figUthS} shows the dependence of the threshold $\Theta_d$ with the density ratio $\rho_p/\rho_f$. It is usually assumed that the Shields number compares directly the horizontal force exerted on a surface grain to its weight, in which case the threshold Shields number could be interpreted as an effective friction coefficient, within a numerical factor. If this was true, $\Theta_d$ would be a constant, independent of $\rho_p/\rho_f$. However, one observes that $\Theta_d$ decreases rapidly with the density ratio.

%%%%%%%%%%%%%%%%%%%%%%%%%%%%%%%%%%%
%%%%%%%%%%%%%%%%%%%%%%%%%%%%%%%%%%%

%%_________________________________________________________
\subsection{Transport layer}
\begin{figure}[t!]
\includegraphics{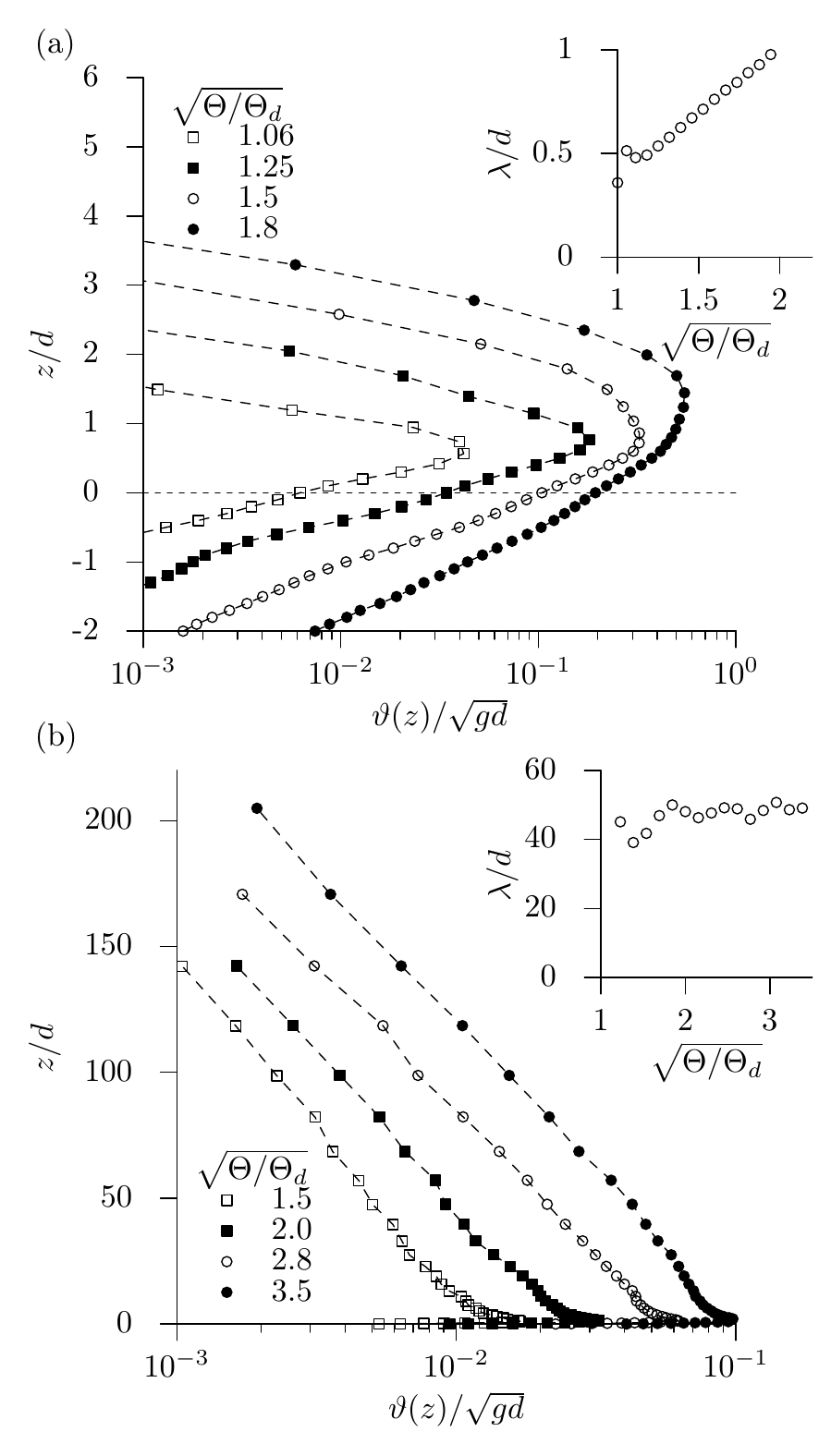}
\caption{Vertical profiles of the sediment flux density $\vartheta(z)$ for different values of the shear velocity ratio $\sqrt{\Theta/\Theta_d}$, in water (a) and air (b). Insets: characteristic transport layer thickness $\lambda$ as function of the shear velocity.}
\label{figQZ}
\end{figure}

Figure~\ref{figQZ} presents the vertical profiles of the flux density, i.e. the flux per unit height $\vartheta(z)$ (such that $q_{\rm sat} \equiv \int \vartheta(z) dz$) for different shear velocities. It shows that bed load and saltation mainly differ by the vertical characteristics of the transport layer. At small density ratios the motion of grains is confined within a thin layer of few grain diameters (Fig.~\ref{figQZ}a). Most of the bed load occurs at about one grain diameter above the static bed and the flux density profile decays symmetrically on both sides of this maximum. By contrast, for large density ratios, grains experience much higher trajectories and the transport layer is much wider. Figure~\ref{figQZ}b shows that the flux density still presents a maximum close to the static bed but decreases exponentially with height.

%%%%%%%%%%%%%%%%%%%%%%%%%%%%%%%%%%%
%%%%%%%%%%%%%%%%%%%%%%%%%%%%%%%%%%%

These qualitative observations can be formalized by defining a characteristic transport layer thickness $\lambda$ from the flux density profile $\vartheta(z)$ as:
\begin{equation}
\lambda = \left( \frac{\int_0^{\infty} (z-\bar z)^2 \, \vartheta(z) dz } { q_{\rm sat} } \right)^{1/2}
\end{equation}
where $\bar z = \frac{1}{ q_{\rm sat}} \int_0^{\infty} \! z \, \vartheta(z) dz$ gives the altitude of the transport layer centre. If the flux profile decreases exponentially, $\lambda$ is the characteristic distance over which this decrease takes place. The variations of $\lambda$ with the shear velocity are presented in the insets of figure~\ref{figQZ}. For underwater bed load, the size of the transport layer is about one grain diameter, gently increasing with the shear velocity from $\lambda \simeq d/2$ to $\lambda \simeq d$. For aeolian saltation the transport layer is indeed wider, with a characteristic size $\lambda \simeq 50 d$ roughly independent of the shear velocity.

\begin{figure}[t!]
\includegraphics{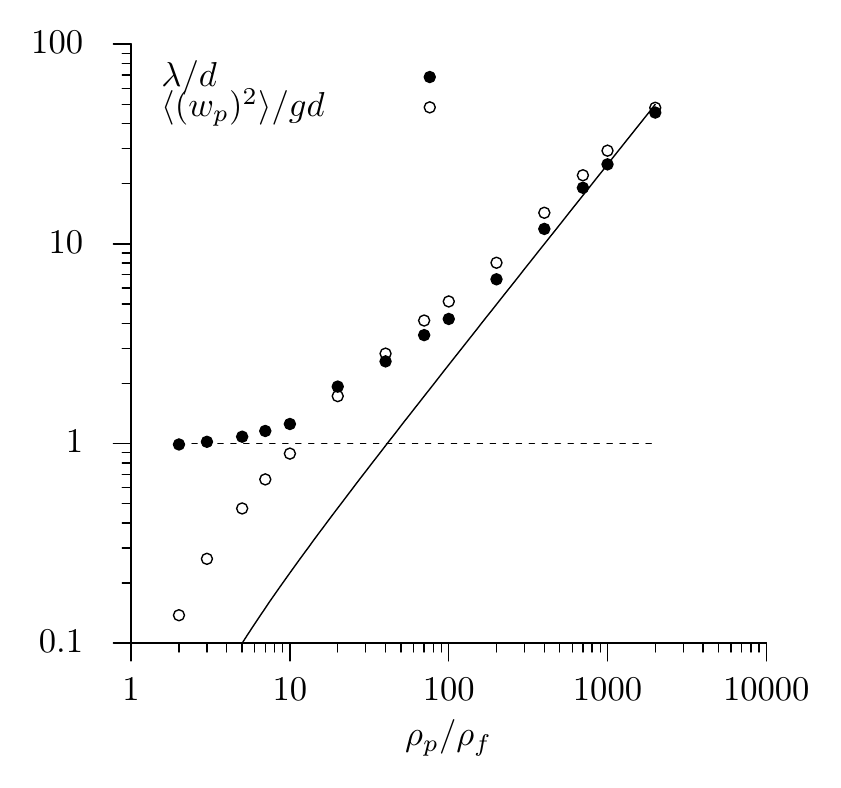}
\caption{Characteristic transport layer thickness $\lambda$ ($\bullet$) as function of the density ratio for $\Theta = 2\Theta_d$. At small density ratios it is limited by the grain size (dashed line), while for large ones it scales as $\rho_p/\rho_f\, d$ (solid line). The averaged vertical energy per grain $\langle (w^p)^2 \rangle/g$ ($\circ$) is also shown to illustrate the dynamical origin of $\lambda$ (see text).}
\label{figLS}
\end{figure}

Figure~\ref{figLS} shows the dependence of the transport layer thickness $\lambda$ with the density ratio. At large density ratios, $\lambda$ is observed to scale with $\frac{\rho_p}{\rho_f}d$, i.e. proportional to the drag length, which emerges when the motion of the grains is dominated by the balance between inertia and hydrodynamical drag. This length is thus expected to control the characteristic hop height and hop length, which naturally leads to wider transport layers for lighter fluids. In sub-aqueous conditions (density ratios of order unity), the transport layer thickness is limited by the grain size $\lambda \sim d$, which is the characteristic length scale for contact forces and geometrical trapping of particles \cite{QADD00,A07}.

%%%%%%%%%%%%%%%%%%%%%%%%%%%%%%%%%%%
%%%%%%%%%%%%%%%%%%%%%%%%%%%%%%%%%%%

The hop height can be estimated from the particle vertical velocity $w^p$ using the ballistic approximation, neglecting the vertical component of the drag force. Under this hypothesis, one expects the hop height to increase like $(w^p)^2/g$. Figure~\ref{figLS} shows the dependence of the average squared vertical velocity $\langle (w^p)^2 \rangle$ on the density ratio. One observes that the transport layer thickness $\lambda$ is effectively determined by the hop length $\langle (w^p)^2 \rangle/g$ for $\rho_p/\rho_f \gtrsim 10$. Below this cross-over value, the transport layer thickness is given by the grain diameter $d$, as trajectories are almost horizontal. The transition from bed load to saltation therefore takes place when the vertical velocities of the particles are sufficiently large for these particles to escape the traps formed by the grains on the static bed. Formally, the criterion of this transition can then be written as $\langle (w^p)^2 \rangle \simeq g d$.

%%_________________________________________________________
\subsection{Grain feedback on the flow and the conditions for saturation}

\begin{figure}[t!]
\includegraphics{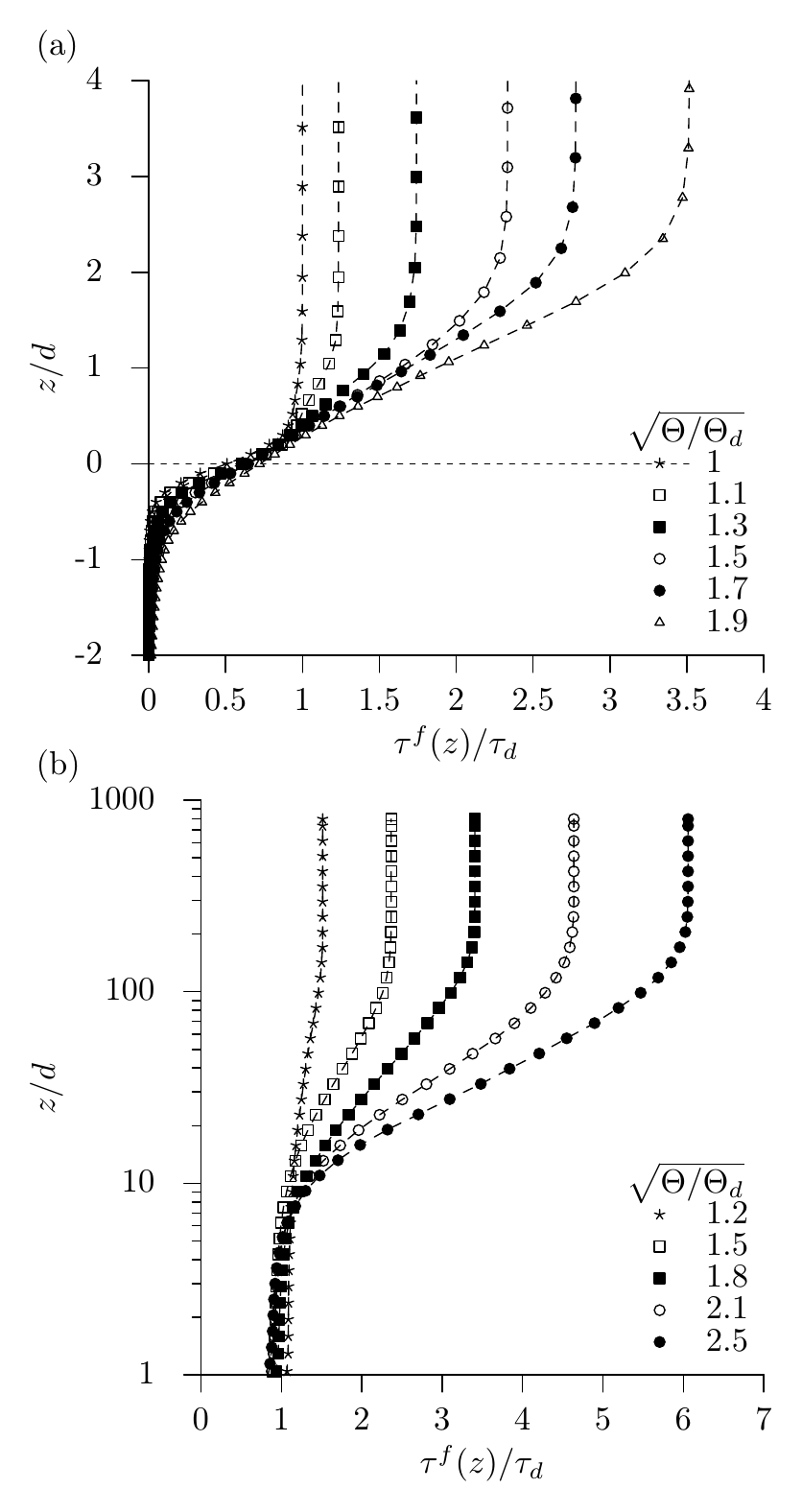}
\caption{Vertical profiles of the fluid borne shear stress $\tau^f(z)$ rescaled by the dynamic threshold $\tau_d = \Theta_d (\rho_p/\rho_f - 1) g d$ for different values of the shear velocity ratio $\sqrt{\Theta/\Theta_d}$, in water (a) and air (b).}
\label{figTauZ}
\end{figure}

Another difference between bed load and saltation is how the grain's feedback on the flow is distributed within the steady state transport layer. Figure \ref{figTauZ} presents the vertical profiles of the fluid shear stress, rescaled by the dynamical threshold $\tau_d = \Theta_d (\rho_p/\rho_f-1) g d$ (as defined by the saturated flux), for different shear velocities. For bed load (Fig.~\ref{figTauZ}a), the different profiles of the fluid shear stress seems to converge to the threshold value very close to the surface ($z=0$). As the dynamical threshold represents the limit shear stress for the fluid to sustain transport, the grain motion is directly driven by the fluid and thus controlled by the excess shear stress $\rho_f u_*^2-\tau_d$. In this transport layer, the fluid momentum decays over few grain sizes, in agreement with the vertical extension of the transport layer. In contrast, the fluid shear stress is below the threshold in the bed ($z<0$) but some (weak) transport still occurs there, which is sustained not by the fluid itself but by the momentum transferred to the surface by grain collisions. 

This general picture is still valid for saltation (Fig.~\ref{figTauZ}b), however now the dynamical threshold is reached much farther from the surface (at $z \simeq 10\,d$) which implies that the kinetic energy of impacting grains is large enough as to sustain the transport below this height. Above it, the transport is driven by the fluid and most of its momentum is dissipated in a much larger layer (comprising tens of grain diameters) again in agreement with the size of the saltation layer. Notice that although this surface sublayer below $10\,d$ contains most of the grains (see Fig.~\ref{fig1}) it still represents a small fraction of the overall transport layer, whose characteristic thickness is $\lambda \simeq 50\,d$.

An important consequence of this distinction in the vertical structure of the grain's feedback is that although for bed load transport is equilibrated when the fluid shear stress reaches its dynamical threshold below the transport layer, this condition is not enough for saltation to equilibrate. For saltation there is a sub-layer where transport is not directly driven by the fluid and thus its equilibration is not dictated by the threshold. There, the properties of grain's collisions become relevant and the equilibrium is described by the conservation of the number of saltating grains i.e. when the number of grains entering the flow exactly balance those grains trapped by the bed. 

%%%%%%%%%%%%%%%%%%%%%%%%%%%%%%%%%%%
%%%%%%%%%%%%%%%%%%%%%%%%%%%%%%%%%%%

%%_________________________________________________________
\section{Interpretation}

%%_________________________________________________________
\subsection{A simple transport model for bedload}
We propose here a simple model of bed load inspired from Bagnold's original ideas \cite{B56}. We hypothesize that moving grains are confined in a thin layer with thickness on the order of $d$. As the average particle vertical velocity is very small, grain hop heights are typically much smaller than $d$ (Fig.~\ref{figLS}), which means that  the vertical motion of the grains can effectively be neglected. The saturated flux can then be decomposed as the product of the number $n$ of transported grains per unit area by the mean grain horizontal velocity $\bar u^p$:
\begin{equation}
q_{\rm sat} = \frac{1}{\phi_b} \, \frac{\pi}{6} d^3 n \bar u^p.
\label{qsatnup}
\end{equation}
In the numerical simulations, we compute $n$ and $\bar u^p$ as
\begin{eqnarray}
n & = & \frac{\left( \sum_p u_p \right)^2}{A \sum_p u_p^2}, \label{defnumn}\\
\bar u^p & = & \frac{\sum_p u_p^2}{\sum_p u_p}. \label{defnumup}
\end{eqnarray}
Notice that these definitions are consistent with the definition of $q_{\rm sat}$ (\ref{defqsatnum}). If all grains were moving at the same velocity, then $n$ and $u^p$ would indeed be respectively the density of moving grains and their velocity.

We can then write the grain born shear stress as proportional to the moving grain density $n$ and to the drag force acting on a grain moving at the average velocity $\bar u^p$ due to a flow at the velocity $u$:
\begin{equation}
\tau^p = n f_d \quad {\rm with} \quad f_d=\frac{\pi}{8}  C_d^\infty  \rho_f \left(u - \bar u^p\right)^2\,d^2.
\end{equation}
For the sake of the argument, we neglect Archimedes contribution as well as the dependence of the drag coefficient on the particle Reynolds number. A key assumption is that grains are in a steady motion, which means that the drag force $f_d$ balances a resistive force due granular friction, collisions with the bed, etc. These different dissipative mechanisms can be modeled as an overall effective friction force characterized by a friction coefficient $\mu_d$:
\begin{equation}
f_d = \frac{\pi}{6} \mu_d (\rho_p - \rho_f) g d^3.
\end{equation}
We can furthermore express the fluid velocity $u_d$ at the transport threshold by assuming that the hydrodynamic drag exerted on a static grain ($u^p=0$) has to overcome a static friction, characterized by a coefficient $\mu_s$:
\begin{equation}
u_d=\sqrt{\frac{4  \mu_s}{3 C_d^\infty} \left(\frac{\rho_p}{\rho_f}-1\right) g d}\;.
\label{ud}
\end{equation}
Combining the above equations it follows that the velocity difference between the grain and the flow is constant:
\begin{equation}
\bar u^p = u - \sqrt{\frac{\mu_d}{\mu_s}} u_d
\end{equation}
We now assume that the transported grains do not disturb the flow. Then, the flow velocity around grains $u$ must be proportional to the shear velocity, so that $u/u_d=\sqrt{\Theta/\Theta_d}$ (see Fig.~\ref{figUZ}a). One therefore deduces:
\begin{equation}
\bar u^p = u_d\;\left(\sqrt{\frac{\Theta}{\Theta_d}} - \sqrt{\frac{\mu_d}{\mu_s}} \right)\, .
\label{upbedload}
\end{equation}
This predicts that the grain velocity does not vanish at the threshold, if friction is lowered during motion ($\mu_d<\mu_s$). The velocity at threshold $u_d (1-\sqrt{\mu_d/\mu_s})$ can be interpreted as the velocity needed by a grain to be extracted from the bed and entrained by the flow.

Saturation is reached when the fluid shear stress equals the transport threshold at the surface of the static bed i.e. when $\tau^p = \rho_f u_*^2 - \tau_d$, with $\tau_d = \Theta_d (\rho_p/\rho_f - 1) g d$ (Fig.~\ref{figTauZ}a). As consequence, the number of transported particles per unit area is solely determined by the excess shear stress:
\begin{equation}
n =  \frac{\rho_f u_*^2 - \tau_d}{f_d} =\frac{\Theta - \Theta_d}{ \frac{\pi}{6} \mu_d d^2} \, .
\label{nbedload}
\end{equation}
Finally, the saturated flux reads:
\begin{equation}
q_{\rm sat} =  \frac{u_d d}{\phi_b \mu_d}\;\left(\Theta - \Theta_d\right)\;\left(\sqrt{\frac{\Theta}{\Theta_d}} - \sqrt{\frac{\mu_d}{\mu_s}} \right).
\label{qsatbedload}
\end{equation}
Inserting the expression (\ref{ud}) of $u_d$, one gets the scaling law for the flux at large $\Theta$:
\begin{equation}
q_{\rm sat} \propto \Theta^{3/2}\;\sqrt{ \left(\frac{\rho_p}{\rho_f}-1\right) g d^3} \, .
\end{equation}
%

%_________________________________________________________
\subsection{A simple transport model for saltation}
We now proceed in a similar manner for the aeolian saltation regime, following ideas initially proposed by Owen (1964) and Ungar \& Haff (1987). In this regime, the motion of the grains is not confined to a thin layer at the surface of the bed. We consider an average grain trajectory, in which the particle takes off from the bed with the horizontal velocity $\bar u^p_\uparrow$, and comes back to it with a velocity $\bar u^p_\downarrow$, after a hop of length $a$. Some momentum is extracted from the wind flow by the grains to perform their jumps, so that the particle shear stress writes
\begin{equation}
\tau^p = \rho_p \phi_b \frac{\bar u^p_\downarrow - \bar u^p_\uparrow}{a} \, q_{\rm sat}.
\end{equation}
Now we use again the decomposition of the saturated flux as the product of the grain density $n$ and the grain velocity $\bar u^p$ (Eq.~\ref{qsatnup}). Saturated transport corresponds to the balance $\tau^p = \rho_f u_*^2 - \tau_d$, so that $n$ still has the same form as in the bed-load case:
\begin{equation}
n = \frac{(\rho_p-\rho_f) g d}{f_d} \left( \Theta - \Theta_d \right),
\label{nsaltation}
\end{equation}
but with a different effective drag force $f_d$, not related to friction anymore but to grain velocities. As the grain hop length can be related to the grain velocity as $a \propto \bar u^p_\uparrow \bar w^p_\uparrow / g$ (ballistic approximation), we can effectively write
\begin{equation}
f_d \propto \frac{\pi}{6} d^3 \rho_p g \, \frac{\bar u^p_{\downarrow}-\bar u^p_{\uparrow}}{\bar w^p_{\uparrow}} \, \frac{\bar u^p}{\bar u^p_{\uparrow}} \, .
\end{equation}

Now, for saltation, steady transport also implies that the number of grains expelled from the bed into the flow exactly balances those trapped by the bed, i.e. a replacement capacity equal to one. Due to the grain feedback on the flow, in contrast with bed load, grains in the transport layer feel a flow independent of the wind strength (see Fig.~\ref{figTauZ}b and Fig.~\ref{figUZ}b). Thus, new moving grains come only from high energy bed collisions. Since the number of ejected grains is a function of the impact energy (or equivalently, of the impact velocity), the mean grain velocity $\bar u^p$ must be constant, independent of the shear velocity, scaling with $u_d$:
\begin{equation}
\bar u^p \propto u_d.
\label{upsaltation}
\end{equation}
From this argument, it follows that all particle surface velocities ($\bar u^p_{\downarrow}$, $\bar u^p_{\uparrow}$, $\bar w^p_{\uparrow}$) also scale with $u_d$, so that $f_d$ is also a constant. Finally, the scaling law followed by the saturated flux becomes,
\begin{equation}
q_{\rm sat} \propto  (1-\rho_f/\rho_p) u_d d \left( \Theta - \Theta_d \right).
\label{qsatsaltation}
\end{equation}
%

%%_________________________________________________________
\subsection{Comparison with simulations}
The above simplifying models suggest a few simple tests to investigate the dynamical mechanisms in the DEM simulation. (i) Is saturation of transport due (or not) to the negative feedback of moving grains on the fluid? (ii) Do we recover the linear relation between the grain density $n$ and the excess Shield number $\Theta - \Theta_d$, whatever the transport regime? (iii) Does the mean grain velocity $\bar u^p$ depend (or not) on the shear velocity?

%%_________________________________________________________
\subsubsection{Grain feedback on the flow}

\begin{figure}[t!]
\includegraphics{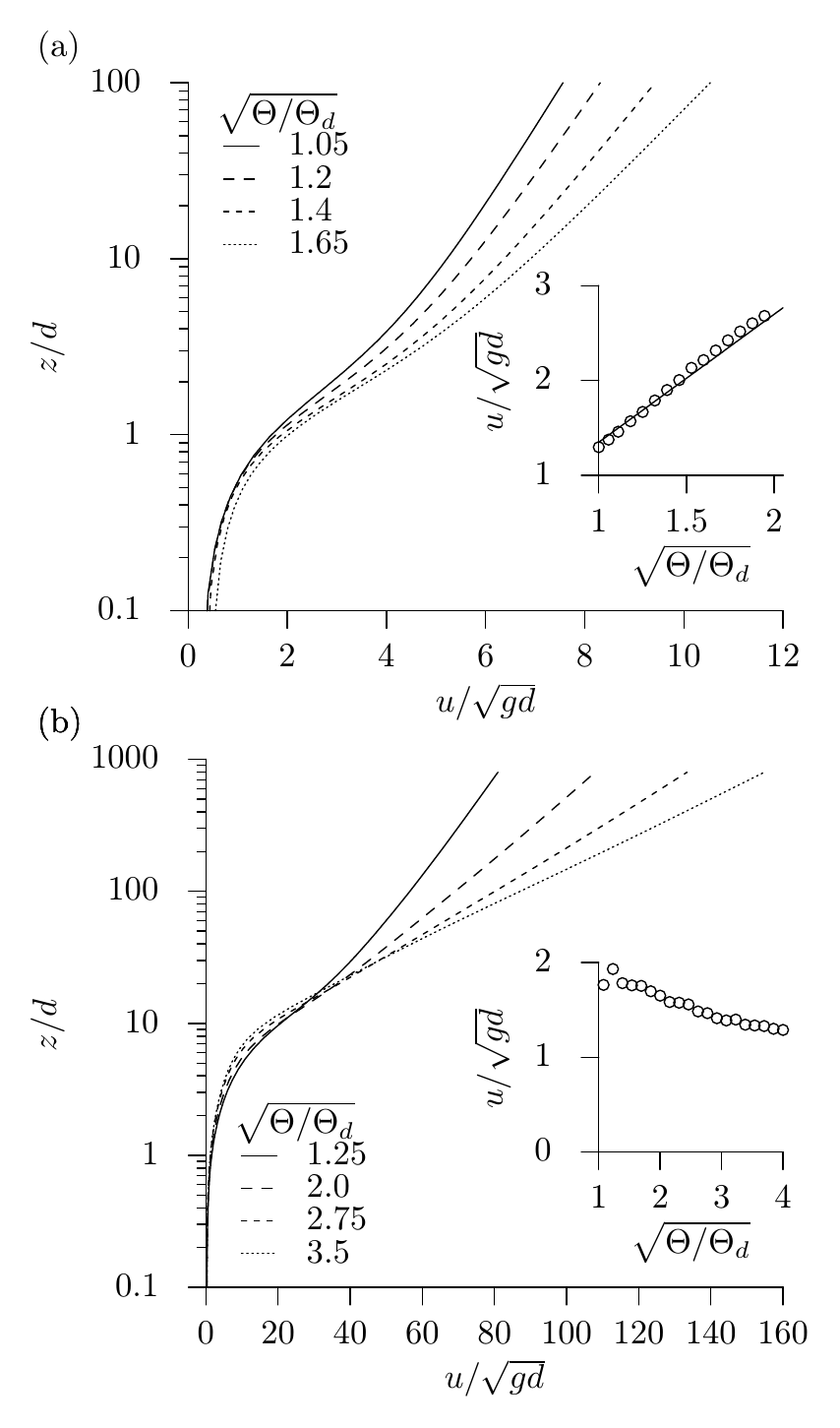}
\caption{Flow velocity vertical profiles at different shear velocity ratios $\sqrt{\Theta/\Theta_d}$ for water (a) and air (b). Insets: velocity at the point of maximum transport ($z \simeq 1d$), as a function of the rescaled shear velocity. The dashed line in the upper inset corresponds to the fit $u \propto u_*$.}
\label{figUZ}
\end{figure}

The information of the feedback of moving grains on the fluid flow is formally encoded in the flow roughness length $z_0$. However, it can first be qualitatively understood from the shape of the fluid shear stress and flow velocity profiles inside the transport layer (Fig.~\ref{figTauZ} and Fig.~\ref{figUZ}, respectively). For bed load, as shown in the inset of Fig.~\ref{figUZ}a, the flow velocity at the point of maximum transport (at about $1d$), increases with the shear velocity. This indicates that the flow is barely disturbed in the transport layer. As shown in Fig.~\ref{figTauZ}a, most of the feedback is actually confined very close to the static bed and thus, most moving grains feel the undisturbed flow that increases with $u_*$. In contrast, for aeolian transport (see Fig.~\ref{figTauZ}b and Fig.~\ref{figUZ}b) the flow velocity is strongly affected by the motion of grains as it becomes almost independent of the wind in the region $z \lesssim 10 d$.

%%%%%%%%%%%%%%%%%%%%%%%%%%%%%%%%%%%
%%%%%%%%%%%%%%%%%%%%%%%%%%%%%%%%%%%

%%%%%%%%%%%%%%%%%%%%%%%%%%%%%%%%%%%
\begin{figure}[t!]
\includegraphics{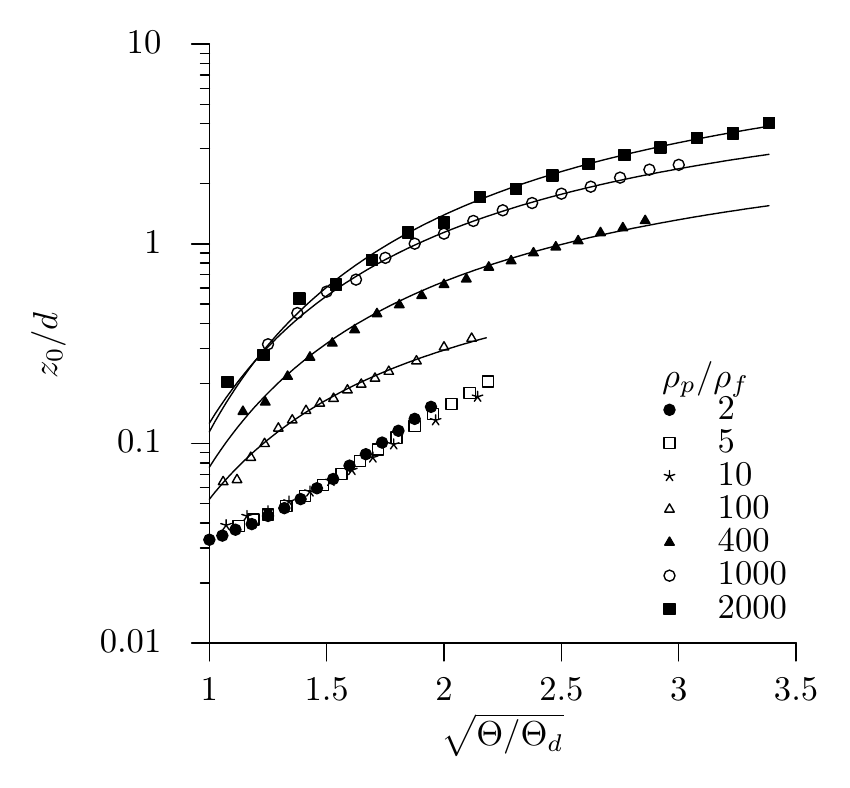}
\caption{Rescaled hydrodynamical roughness length as function of the shear velocity for different density ratios. Solid lines are the predictions based on the focal point assumption (Eq.~\ref{z0focalpoint}).}
\label{figZ0}
\end{figure}
%%%%%%%%%%%%%%%%%%%%%%%%%%%%%%%%%%%

The data of the hydrodynamical roughness length $z_0$ show a similar picture (Fig.~\ref{figZ0}). In the saltation regime the roughness length increases with the shear velocity as a result of grain feedback, which can be modeled from the existence of a focal point where $u=U_f$ at $z=H_f$ independently of $u_*$ (Fig.~\ref{figUZ}b), and above which the flow velocity recovers its log profile $u=u_*/\kappa \, \ln (z/z_0)$ \cite{LD04,RS08,CDOVCJPR09}. This gives
\begin{equation}
z_0 \simeq H_f \exp{(-\kappa U_f/u_*)}.
\label{z0focalpoint}
\end{equation}
This expression reproduces well the increase of $z_0$ for stronger winds, when the density ratio $\rho_p/\rho_f $ is large enough (Fig.~\ref{figZ0}). Typically below $\rho_p/\rho_f \simeq 10$, Eq.~\ref{z0focalpoint} does not reproduce the data anymore. This is consistent with the absence of a focal point in the bed load regime (Fig.~\ref{figUZ}a). Also, in the small $\rho_p/\rho_f $ limit, the roughness length remains very small (substantially smaller than $d$).

%%_________________________________________________________
\subsubsection{Number of transported grains and average grain velocity}

\begin{figure}[t!]
\includegraphics{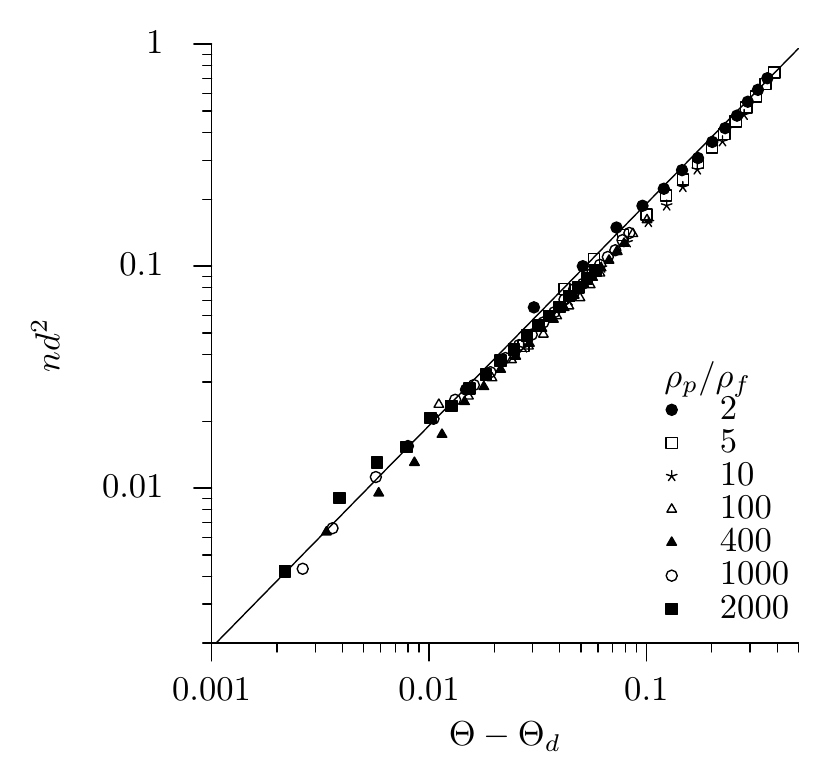}
\caption{Linear relation between the rescaled number of transported grains per unit area and the rescaled excess of shear stress for different density ratios.}
\label{figNU}
\end{figure}

From expressions (\ref{defnumn}) and (\ref{defnumup}), we can compute the number of transported grains per unit area and the mean grain horizontal velocity as a function of the shear velocity of the flow.
Figure \ref{figNU} shows a linear relation between $n$ and $\Theta - \Theta_d$ for both bed load and saltation. This is consistent with the predictions of the above models. Interestingly, the friction coefficient $\mu_d$, defined from the proportionality factor (see Eq.~\ref{nbedload}), has the same value $\simeq 1$ in both cases. This suggests that dissipation due to collisions of the moving grains with the bed plays the same role in both transport regimes.

%%%%%%%%%%%%%%%%%%%%%%%%%%%%%%%%%%%
%%%%%%%%%%%%%%%%%%%%%%%%%%%%%%%%%%%

%%%%%%%%%%%%%%%%%%%%%%%%%%%%%%%%%%%
\begin{figure}[t!]
\includegraphics{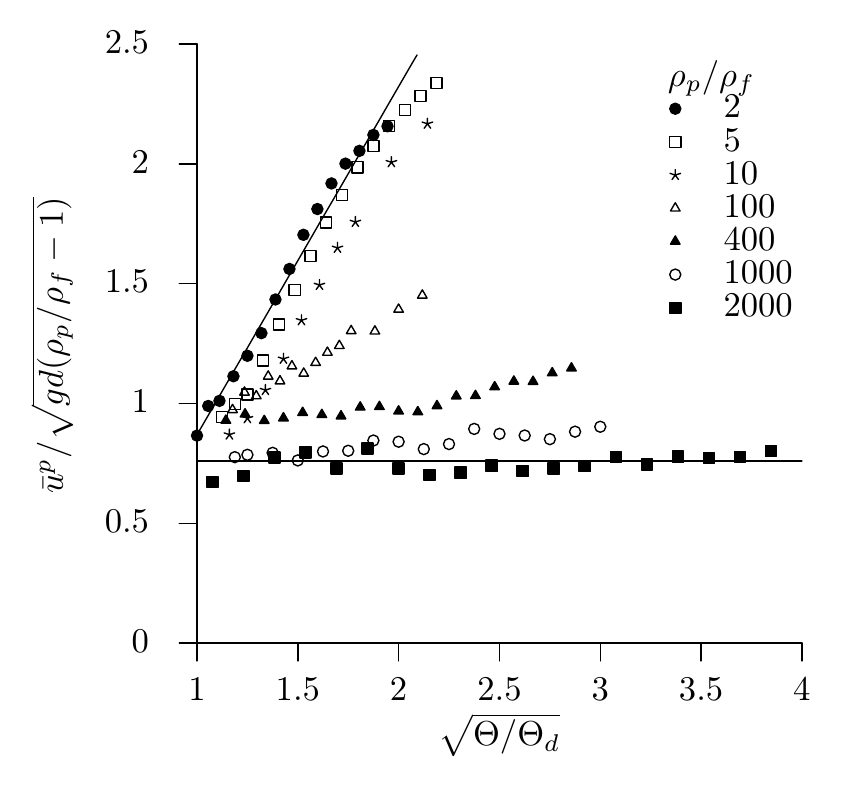}
\caption{Rescaled mean grain velocity as function of the rescaled the shear velocity for different density ratios. Full lines show the analytical prediction given in the text for the two limiting cases: water (Eq.~\ref{upbedload}) and air (Eq.~\ref{upsaltation}).}
\label{figUU}
\end{figure}
%%%%%%%%%%%%%%%%%%%%%%%%%%%%%%%%%%%

The dependence of the mean grain velocity $\bar u^p$ is also fully consistent with the picture emerging from the simple models. As shows in Fig.~\ref{figUU}, $\bar u^p$ increases linearly with $\sqrt{\Theta/\Theta_d}$ for bed load (Eq.~\ref{upbedload}) while it remains roughly constant for aeolian saltation (Eq.~\ref{upsaltation}). Interestingly, the different curves shown in Fig.~\ref{figUU} cross at $\Theta \simeq \Theta_d$. In other words, the grain velocity at the transport threshold scale on $\sqrt{gd (\rho_p/\rho_f - 1)}$, with a prefactor slightly smaller than unity, whatever the transport regime. This common behavior between bed-load and saltation results from the fact that the negative feedback of transport on the flow disappears at the threshold, as $n$ vanishes.

Fitting the grain density and the mean grain velocity to the simple model of bed load, one can extract the effective friction coefficients $\mu_d$ and $\mu_s$. The static friction coefficient $\mu_s$ turns out to be $\simeq 6$ times larger than the dynamical friction coefficient $\mu_d$. This means that the  motion is lubricated by the fluid once the grains are entrained. Therefore, the grain velocity at the threshold remains finite (but $n$ vanishes).

%%_________________________________________________________
\section{Saturation transient}
Beyond the properties of steady and homogeneous transport, we address in this section the time and length scales involved in the relaxation of the sediment flux toward its saturated value, which are relevant in the context of pattern formation \cite{ACD02a,ACD02b,HDA02,CA06,FCA10,ACDDF12}. We further emphasize the difference between the saturation time and the exchange time.

%%_________________________________________________________
\subsection{Saturation length and time}
Whatever the transport regime, the saturation transient is controlled by two mechanisms. On the one hand, bed erosion or deposition must take place to adapt the number of transported grains to the flow velocity. On the other hand, grains must be accelerated by the flow to their asymptotic velocity. 

We have addressed the case of saltation in a series of articles, starting from a controversy between us \cite{ACD02b,DH06,PDH07} and resolving it \cite{DCA11}. In summary, the horizontal acceleration of a grain entrained by the wind  is governed by the equation of motion:
\begin{equation}
\frac{{\rm d} u^p}{{\rm d}t} = \frac{3}{4} \frac{C_d^\infty \rho_f}{\rho_p d} (u-u^p)^2.
\end{equation}
Contrarily to bed load, in the saltation regime, dissipation only takes place during collisions and not through a permanent friction on the static bed. The only length scale in this equation is the so-called drag length $\frac{\rho_p}{\rho_f}  d$. As a consequence, the relaxation of the particle velocity to the fluid velocity occurs over a length which varies as
\begin{equation}
L_{\rm sat} \propto \frac{\rho_p}{C_d^\infty \rho_f} \,  d,
\label{eq:lsatsal}
\end{equation}
independently of the wind speed, with a proportionality factor that depends on the restitution coefficient $e$ \citep{A04}. Except in the vicinity of the transport threshold, the  length over which the number of grains transported relaxes to its saturated state is much shorter than the drag length --~it decays as $1/u_*^2$. Therefore, the overall saturation length is proportional to the drag length, as confirmed by direct measurements \cite{ACP10}.

The case of bed-load is still under debate \cite{C06,FCA10,LMC10}. We derive here the saturation time and the saturation length in the simple bed-load model detailed above. As the moving grains form a surface layer of thickness $d$, the number of moving particles per unit area adapts immediately to a change of shear velocity. By contrast, the grain velocity relaxes to its asymptotic value with a characteristic time. This is what gives the saturation time. Neglecting the dependence of the drag coefficient on the particle Reynolds number, the horizontal component of the grain equation of motion reads:
\begin{equation}
\frac{du^p}{dt}=\frac{3 C_d^\infty \rho_f}{4 \rho_p d} \left[(u-u^p)^2-\frac{\mu_d}{\mu_s}u_d^2\right].
\label{accgrainbedload}
\end{equation}
Linearising this equation around the asymptotic value, we obtain the following expression for the saturation time:
\begin{equation}
T_{\rm sat}=\sqrt{\frac{\mu_s}{\mu_d}}\;\frac{2 \rho_p d}{3 C_d^\infty \rho_f u_d} \,.
\end{equation}
Using expression (\ref{ud}) for $u_d$ and typical values for the various parameters, we get $T_{\rm sat}$ on the order of few $\sqrt{d/g}$. The saturation length is then the length over which the grain moves during $T_{\rm sat}$ at velocity $\bar u^p$:
\begin{equation}
L_{\rm sat} = \frac{2}{3} \, \frac{\rho_p d}{C_d^\infty \rho_f } \left(\sqrt{\frac{\mu_s}{\mu_d}} \frac{u}{u_d} -1\right).
\end{equation}
Inserting again typical numbers in this expression, we get, for $u$ close to $u_d$, a value for $L_{\rm sat}$ on the order of few grain diameters.

%%_________________________________________________________
\subsection{Exchange time vs saturation time}

An important problem that cannot be tackled using the simple transport models presented here (or any Eulerian continuous model) is the exchange between the mobile and the static phases. Such models do not aim to describe the Lagrangian paths of individual grains. In particular, recent studies have focused on the characteristic time a given grain spends in the transport layer before being trapped by the bed \cite{C06,FCA10,LMC10}. This time, noted  $T_{\rm ex}$ hereafter, is either called the deposition time or the exchange time. It is relevant in geology as it reflects the time scale associated with storage and reworking of sediments. The exchange time should not be confused with the saturation time. Imagine for instance the case where all the grains in the transport layer would move with a uniform and perfectly horizontal velocity. Then there would be no exchange with the static phase and $T_{\rm ex}$ would be infinite, although transport could reach saturation after a very short time. Despite this conceptual difference, the formalism proposed in \cite{C06} leads to an identity between the exchange time and the saturation time.

\begin{figure}[t!]
\includegraphics{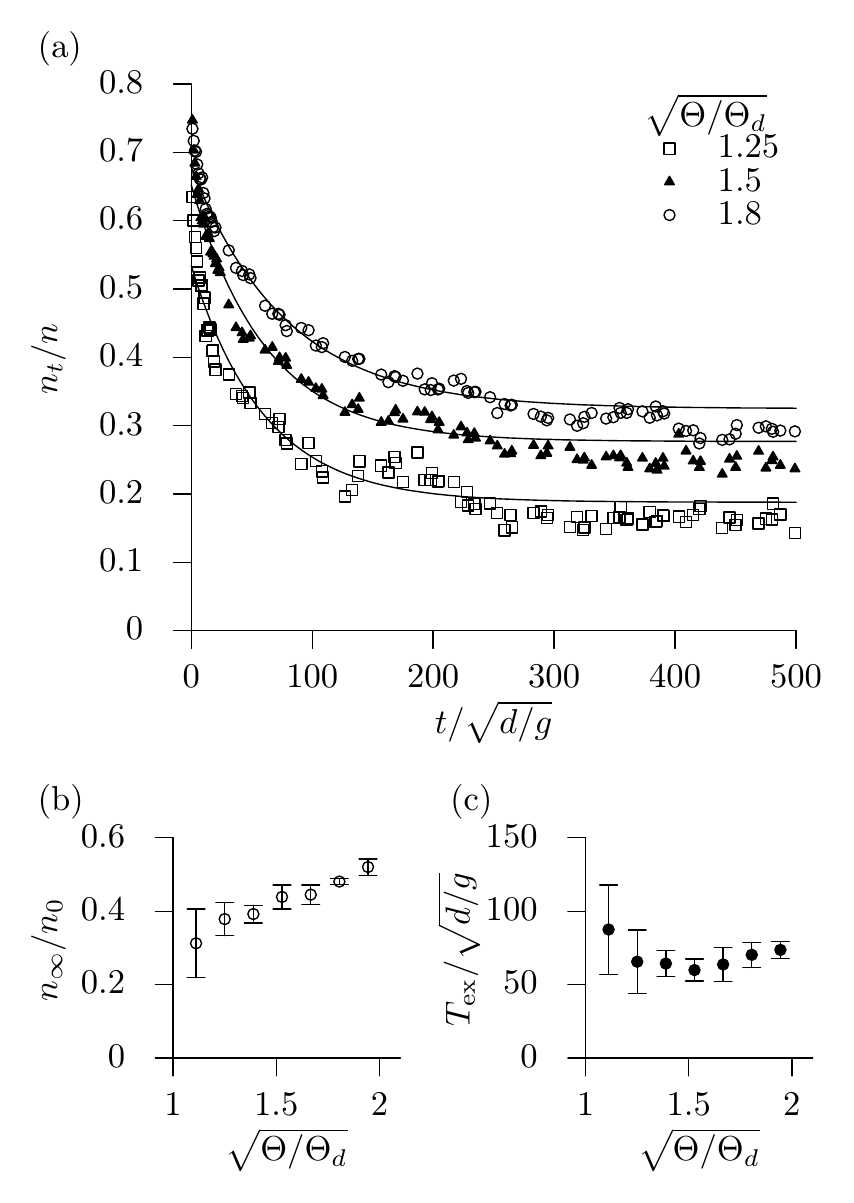}
\caption{(a) Time decay of the fraction of grains with an initial velocity above $\sqrt{g d}/2$ that remain in the transport layer at time $t$, for different shear velocities. The solid line is the exponential fit (Eq.~\ref{ExpoDecaynt}). Panels (b) and (c) respectively show the ratio of the asymptotic to initial value and the characteristic exchange time, as a function of the rescaled shear velocity.}
\label{figExt}
\end{figure}

Using our granular based transport simulations, we address this issue for bed load ($\rho_p/\rho_f = 2$) by tracking all grains with velocities above a certain value at $t=0$. For the sake of this discussion, we have chosen this value to be $\sqrt{g d}/2$, which  allows us to determine the grains inside the transport layer at the initial time. Noting this particle ensemble $\mathcal{E}$, we define the density $n_t$ of moving grains at time $t=0$ that remains in the transport layer after a time $t$:
\begin{equation}
n_t = \frac{\left( \sum_{p \in \mathcal{E}} u_p \right)^2}{A \sum_{p \in \mathcal{E}} u_p^2} \, ,
\end{equation}
The time evolution of $n_t$ in our numerical simulations is displayed in Fig.~\ref{figExt}a. It can be seen that $n_t(t)$ follows an exponential relaxation with time
\begin{equation}
n_t (t) = (n_0 - n_{\infty}) \exp{(-t/T_{\rm ex})}  + n_{\infty},
\label{ExpoDecaynt}
\end{equation}
where $n_0$ and $n_{\infty}$ are the initial and asymptotic values, respectively. As the grains that no longer move have been exchanged with the static phase, the relaxation time of $n_t$ is by definition the exchange time $T_{\rm ex}$.

%%%%%%%%%%%%%%%%%%%%%%%%%%%%%%%%%%%
%%%%%%%%%%%%%%%%%%%%%%%%%%%%%%%%%%%

When analyzed for different shear velocities, the fraction of grains re-entrained in the flow after being trapped by the bed, which is given by the ratio $n_{\infty}/n_0$ (Fig.~\ref{figExt}b), depends weakly on $\sqrt{\Theta}$. However, it seems to tend to zero at the threshold, which is a reasonable limit as all transported grains should be eventually trapped by the surface and replaced by new ones. The exchange time is also roughly constant, with a mean value $T_{\rm ex} \simeq 70 \sqrt{d/g}$ (Fig.~\ref{figExt}c). This time is larger at least by one order of magnitude than the saturation time. This means that exchange between the bed and the transport layer is not the dominant mechanism for the relaxation of the sediment flux towards saturation.

%%_________________________________________________________
\section{Conclusions}

The aim of this paper was to present a novel numerical approach for sediment transport based on a discrete element method (DEM) for particles coupled to a continuum Reynolds averaged description of hydrodynamics. We have studied the effect of the grain to fluid density ratio $\rho_p/\rho_f $ and showed that we can reproduce both (sub-aqueous) bed load at $\rho_p/\rho_f $ close to unity, where transport occurs in a thin layer at the surface of the static bed, and (aeolian) saltation at large $\rho_p/\rho_f $, where the transport layer is wider and more dilute.

We have studied the mechanisms controlling steady, or saturated transport. In the bed load case, saturation is reached when the fluid borne shear stress at the interface between the mobile grains and the static grains is reduced to its threshold value. The number of grains transported per surface unit is therefore limited by the available momentum at the bed surface. However, the fluid velocity in the transport layer remains almost undisturbed so that the mean grain velocity scales with the shear velocity $u_*$. In the saltation case, particles in motion are able to eject others when they collide with the static bed, and saturation is reached when one grain is statistically replaced by exactly another one after collision. As a consequence, the mean grain velocity scales on the shear velocity threshold $u_d$, independently of $u_*$. This provides evidence for a strong negative feedback of the moving grains on the flow within the transport layer, where the wind velocity is reduced. In both bed load and saltation regimes, the number of grains transported per unit area is found proportional to the distance to threshold $\Theta-\Theta_d$, with an identical prefactor on the order of $1/d^2$.

We have systematically varied the density ratio in order to reveal the transition between these two transport regimes. This is also relevant for sediment transport in extraterrestrial atmospheres (Mars, Venus and Titan) \cite{CA06,APAH08a,APAH08b,K10}. We have shown that the properties of bed load transport are observed when $\rho_p/\rho_f \lesssim 10$), whereas those of aeolian saltation are well established when $\rho_p/\rho_f $ is larger than few hundreds. Finally, we have discussed the saturation transient of sediment transport. Based on the mechanisms identified in the steady case, we have derived expressions for the saturation time and length in the two regimes. In the bed load case, we have also shown that the exchange time, which reflects the time scale associated to exchange of particles between the mobile and static phases is an order of magnitude larger than the saturation time.

This study could be continued in different directions. First, it would be interesting to look at the case where the bed is non erodible. This situation has been experimentally investigated in the aeolian regime \cite{HVDM11}, showing a much wider transport layer $\lambda$, and new scaling laws for $\lambda$, the roughness $z_0$ and the flux $q_{\rm sat}$ as a function of $u_*$. Further work should also be done to perform direct measurements of $L_{\rm sat}$ and $T_{\rm sat}$. However, the study of inhomogeneous or unsteady situations requires a finer implementation of the model, especially for averaging procedures. Finally, one shall take into account the turbulent fluctuations and to address the case of suspended transport \cite{CCA11}.

%____________________________________________________________________________
\begin{acknowledgments}
We thank Stefan Luding for kindly allowing us to use his MD code to simulate the granular system. We are also grateful to Jennifer Johnson for a careful reading of the manuscript. We thank Fran\c cois Charru for useful discussions. This work has benefited from the financial support of the Agence Nationale de la Recherche, grant `Zephyr' ($\#$ERCS07\underline{\ }18).
\end{acknowledgments}

%________________________________________________________________________

%________________________________________________________________________
\end{document}